\PassOptionsToPackage{pdfpagelabels=false}{hyperref}
\documentclass{aa}
\usepackage[varg]{txfonts}
\pdfoutput=1
\usepackage{graphicx}
\usepackage{newtxtext,newtxmath}
\usepackage{xspace}
\usepackage{bm}
\RequirePackage[colorlinks=true,linkcolor=blue,citecolor=blue,urlcolor=blue]{hyperref}

\usepackage{caption} 
\usepackage{amssymb}

\newcommand{\msun}{\,{M$_{\odot}$}\xspace}

\newcommand{\Lya}{\ifmmode{\mathrm{Ly}\alpha}\else Ly$\alpha$\xspace\fi}

\begin{document}

\title{Introducing cosmosTNG: simulating galaxy formation\\ with constrained realizations of the COSMOS field}
\titlerunning{cosmosTNG}

\author{Chris Byrohl\inst{1}\thanks{E-mail: chris.byrohl@uni-heidelberg.de}
\and Dylan Nelson\inst{1}
\and Benjamin Horowitz\inst{2,3}
\and Khee-Gan Lee\inst{2,3}
\and Annalisa Pillepich\inst{4}
}

\institute{Universität Heidelberg, Institut für Theoretische Astrophysik, ZAH, Albert-Ueberle-Str. 2, 69120 Heidelberg, Germany \label{1}
\and Kavli IPMU (WPI), UTIAS, The University of Tokyo, Kashiwa, Chiba 277-8583, Japan \label{2}
\and Center for Data Driven Discovery, Kavli IPMU (WPI), UTIAS, The University of Tokyo, Kashiwa, Chiba 277-8583, Japan \label{3}
\and Max-Planck-Institut f\"{u}r Astronomie, K\"{o}nigstuhl 17, 69117 Heidelberg, Germany \label{4}
}

\date{}

\abstract{
We introduce the new cosmological simulation project \textbf{cosmosTNG}, a first-of-its-kind suite of constrained galaxy formation simulations for the universe at Cosmic Noon ($z\sim 2$). cosmosTNG simulates a $0.2$\,deg$^2$ patch of the COSMOS field at $z \simeq 2.0-2.2$ using an initial density field inferred from galaxy redshift surveys and the CLAMATO \Lya forest tomography survey, reconstructed by the TARDIS algorithm. We evolve eight different realizations of this volume to capture small-scale variations. All runs use the IllustrisTNG galaxy formation model with a baryonic mass resolution of $10^6$\,M$_\odot$, equal to TNG100-1. In this initial study, we demonstrate qualitative agreement between the evolved large-scale structure and the spatial distribution of observed galaxy populations in COSMOS, emphasizing the zFIRE protocluster region. We then compare the statistical properties and scaling relations of the galaxy population, covering stellar, gaseous, and supermassive black hole (SMBH) components, between cosmosTNG, observations in COSMOS, and $z \sim 2$ observational data in general. We find that galaxy quenching and environmental effects in COSMOS are modulated by its specific large-scale structure, particularly the collapsing protoclusters in the region. With respect to a random region of the universe, the abundance of high-mass galaxies is larger, and the quenched fraction of galaxies is significant lower at fixed mass. This suggests an accelerated growth of stellar mass, as reflected in a higher cosmic star formation rate density, due to the unique assembly histories of galaxies in the simulated COSMOS subvolume. The cosmosTNG suite will be a valuable tool for studying galaxy formation at cosmic noon, particularly when interpreting extragalactic observations with HST, JWST, and other large multi-wavelength survey programs of the COSMOS field.
}


\maketitle

\section{Introduction}

Initial density fluctuation of dark matter quickly grow after the Big Bang, forming a cosmic web of filaments, sheets and voids~\citep{Bond96}. Baryonic matter soon follows the gravitational collapse of dark matter, enabling the formation of galaxies within halos.
This large-scale environment plays a crucial role in shaping the properties of galaxies, particularly during the epoch of cosmic noon, the peak of star formation and quasar activity at $z\sim 2-3$ roughly 10 billion years ago, as galaxies rapidly evolve through cold gas streams and mergers~\citep{Lacey93, Dekel09}. Recent and upcoming surveys provide a wealth of data during this epoch. With the launch of the James Webb Space Telescope (JWST), several programs are providing an unprecedented view on high-redshift galaxies at cosmic noon and beyond~\citep{Dunlop21, Treu22, Finkelstein23, Casey23}. Similarly, near-infrared multi-object spectrographs such as future VLT-MOONS and Subaru-PFS~\citep{Maiolino20, Greene22} will supplement these space-based efforts at Cosmic Noon.

Galaxy formation simulations have become an important tool for the interpretation of such observations~\citep{Hopkins18, Vogelsberger20a}. In particular, simulations of large cosmological volumes can now reproduce many observed key properties and scaling relations for statistical samples of galaxies~\citep{Pillepich18a, Lee21, Schaye23}. Conventional cosmological simulations are initialized with a random Gaussian field matching a given power spectrum $P(k)$. This approach allows a statistical comparison to observed data, but does not reproduce the large-scale structure of any given region of the real universe. Typical simulation volumes ($\sim 100^3$\,cMpc$^3$) are subject to cosmic variance effects, particularly at the high mass end. While tuned initial conditions can be used to draw representative volume~\citep[e.g. in TNG, see][]{Vogelsberger14}, caution is needed when comparing to galaxies from limited survey volumes.

To overcome this limitation, `constrained' simulations enable a more nuanced comparison of galaxies in their large-scale environment with observed data. Such simulations utilize existing observations to derive their initial conditions, such that they reproduce the observed large-scale structure at late times. Primarily, this technique is employed in the context of the Local universe, including the ELUCID~\citep{Wang16}, CLONE~\citep{Sorce18a}, and SLOW~\citep{Dolag23} simulations. The CLUES~\citep{Gottloeber10}, HESTIA~\citep{Libeskind20} and SIBELIUS~\citep{Sawala22,McAlpine22} simulations apply these approaches to the local universe ($z\lesssim 0.1$).
Constrained realizations have mostly been run as dark-matter only simulations, but follow-ups and zoom-in versions with hydrodynamics and galaxy formation physics exist \citep[in addition to above see e.g.][]{Yepes14,Li22,Luo24}.

Recently, efforts have been made to extend constrained simulations beyond the local universe. Notably, the COSTCO dark-matter only simulation suite~\citep{Ata22} uses a Bayesian inference algorithm \citep{Kitaura21} with Hamiltonian Monte Carlo Sampling for estimating the initial conditions and underlying matter density field from observed $z\sim 2-3$ galaxies within the COSMOS field \citep{Ata21}. Subsequent N-body re-simulations of these initial conditions reveal the emergence and fate of massive halos and protocluster regions therein \citep{Ata22}.

\begin{figure}
    \includegraphics[width=0.95\columnwidth]{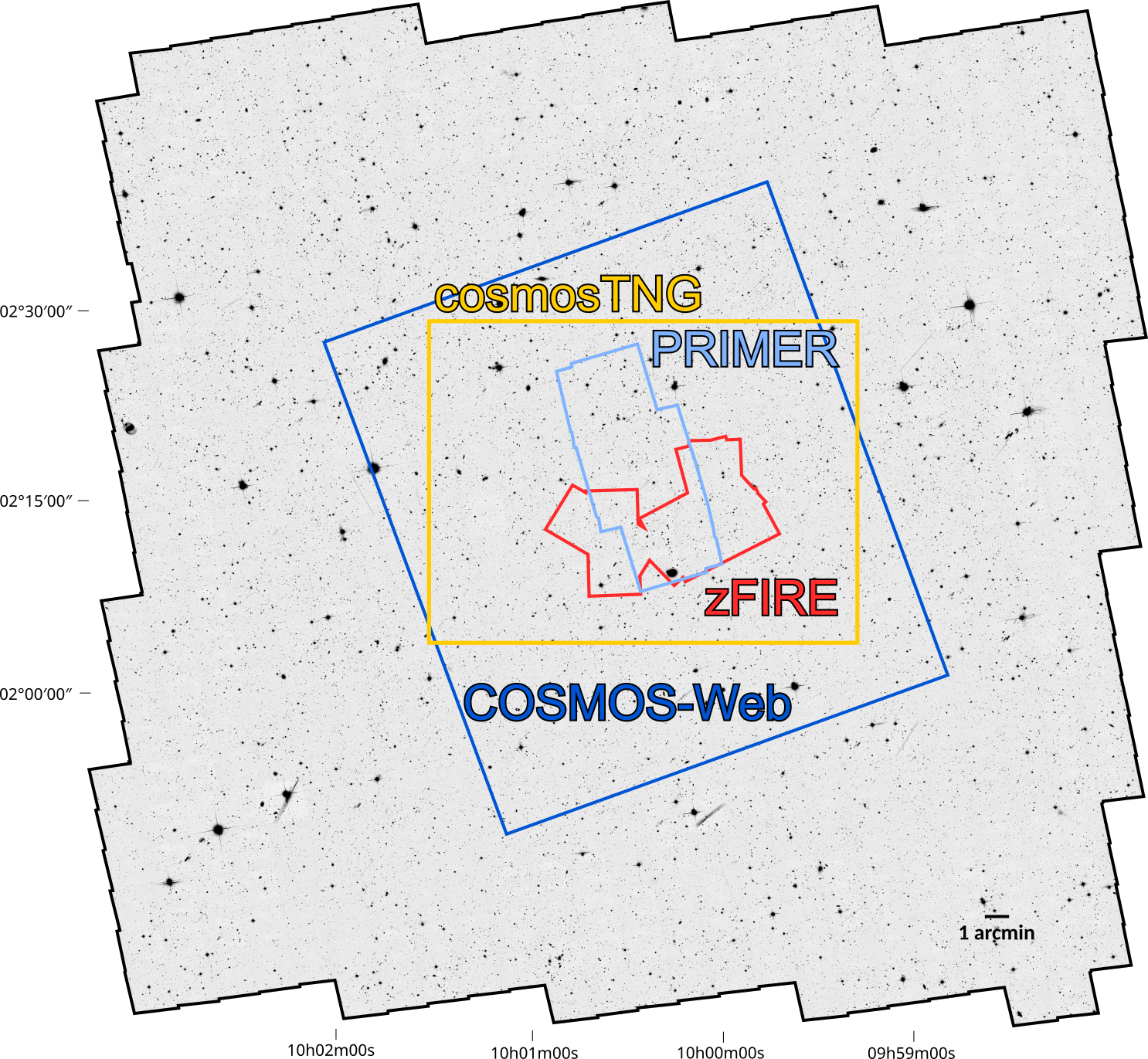}
    \caption{The COSMOS field and the cosmosTNG simulation. The background shows HST-ACS F814W imaging of the field. We show the footprint of cosmosTNG/CLAMATO in the field, along with the zFIRE, COSMOS-Web, and PRIMER survey footprints.}
    \label{fig:cosmosfield}
\end{figure}

The COSMOS field~\citep{Scoville07a} is one of the deepest and most well-studied extragalactic fields. Following its initial deep imaging with ACS on the Hubble Space Telescope, COSMOS has been observed through various programs across the electromagnetic spectrum, including X-ray observations with XMM/Newton and Chandra, optical with Subaru and the VLT, infrared with Spitzer and Herschel, sub-mm with ALMA, and radio with VLA and VLBA. In addition, several large spectroscopic programs with Subaru, Keck and VLT provide redshifts for tens of thousands of galaxies~\citep{Lilly07, LeFevre15, Kriek15, Nanayakkara16}. Much of this data has been compiled in various catalogs~\citep{Capak07, Ilbert09, Laigle16, Weaver22} including more than a million sources with photometric redshifts. The COSMOS field has been used for a wide range of studies, including the relation between galaxies and their host dark matter halos~\citep{McCracken15, Legrand19}, star-formation regulation~\citep{Wang22} and quenching~\citep{Peng10, Moustakas13, Edward24}, and correlations with, and reconstruction of, the large-scale structure~\citep{Scoville13, Darvish17, Laigle18}. 
Uniquely among the various observational deep fields, COSMOS moves beyond a narrow `pencil-beam' geometry and covers a non-trivial footprint on the sky equivalent to $\sim 100\,$cMpc transverse scales at $z>2$.

Importantly, COSMOS has also been targeted for \Lya forest tomography~\citep{Lee14}. It is currently surveyed by JWST as part of various legacy programs such as the COSMOS-Web, PRIMER , Blue Jay and COSMOS-3D programs~\citep{Dunlop21,Casey23,Belli24,Kakiichi24}, which will expand the available data on high-redshift galaxies and their environments in this field. Given this wealth of information for both reconstruction and comparison, the COSMOS field is a prime candidate for further investigation with constrained simulations.

Here, we introduce the new cosmosTNG simulation suite, a set of constrained simulations in the COSMOS field at cosmic noon. cosmosTNG is the first constrained high-redshift cosmological galaxy formation simulation ($z\gtrsim 2$), enabling the study of galaxy formation in the actual large-scale environments that are characterized in observations. Figure~\ref{fig:cosmosfield} shows the COSMOS field and the cosmosTNG footprint, as well as the overlapping, targeted regions by th zFIRE, PRIMER and COSMOS-Web survey footprints.
The simulations are based on the IllustrisTNG model~\citep{Weinberger17, Pillepich18a} and are designed to reproduce the $z=2-2.5$ COSMOS field. The TNG model has been calibrated exclusively on $z=0$ galaxy properties such as the stellar mass function and stellar-to-halo-mass relation, and the cosmic star formation rate density versus time~\citep{Pillepich18a}. This makes galaxy properties and galaxy populations at high redshift particularly predictive in nature. As such, cosmosTNG also enables a unique test of the TNG model in this regime.

This paper is structured as follows. In Section~\ref{sec:methods}, we describe the initial conditions and simulation setup, the TNG galaxy formation model, and the cosmosTNG suite. In Section~\ref{sec:results}, we present the initial results from cosmosTNG, including the reconstructed large-scale structure and a number of galaxy population statistics and scaling relations in comparison to data. We conclude in Section~\ref{sec:conclusions} with a summary of the results and a future outlook.

\section{Methods}
\label{sec:methods}

\subsection{Initial Conditions and Simulation Setup}
\label{sec:simsetup}

\begin{figure}
    \includegraphics[width=0.95\columnwidth]{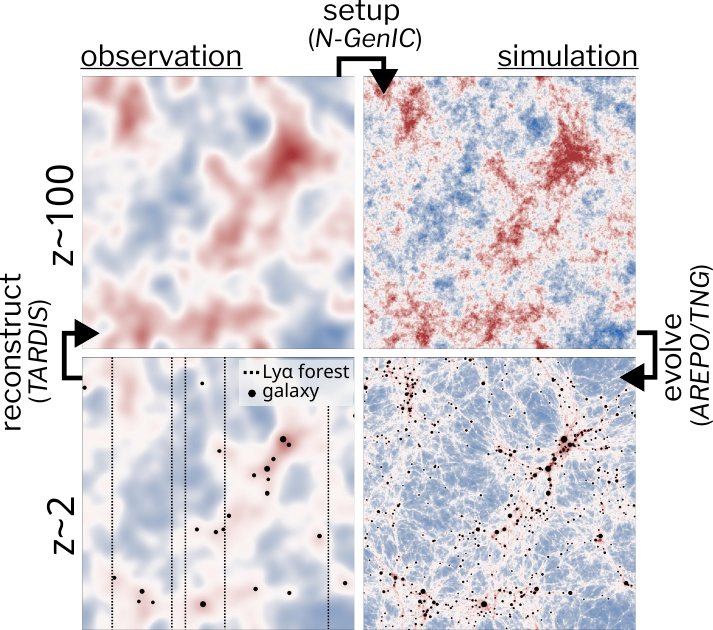}
    \caption{Schematic of the method. We reconstruct the initial $z \sim 100$ density field from observations of zCOSMOS galaxies and CLAMATO \Lya forest measurements at $z \sim 2$ using the TARDIS pipeline. We then generate constrained initial conditions using our modified version of N-GenIC, and finally run the full hydrodynamical simulation using the \textsc{AREPO} code.}
    \label{fig:methods}
\end{figure}

In the following, we describe our methods for constructing and running constrained simulations. The overall method is summarized in Figure~\ref{fig:methods}, showing the key steps of reconstruction, initial conditions setup, and running of the hydrodynamical simulation.

\begin{figure}
    \includegraphics[width=0.95\columnwidth]{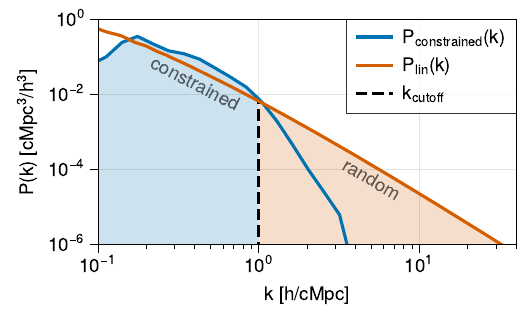}
    \caption{Power spectrum of the simulation volume at its initial condition ($z\sim 127$). Blue shows the power of the constrained field (large scales) in the $34\times 28\times 450$\,cMpc$^3$/h$^3$ volume. Orange shows the theoretically expected linear power. The shaded region show the composite realization for the cosmosTNG simulations: modes are taken from the constrained field up to k$_\mathrm{cutoff}=1$\,h/cMpc above which we randomly draw modes according to linear theory.}
    \label{fig:power_constrained}
\end{figure}

The initial conditions are derived from observations of the COSMOS Lyman-alpha Mapping And Tomography Observations (CLAMATO) Survey~\citep{Lee14, Lee18, Horowitz21}, supplemented by spectroscopic galaxy samples added in the TARDIS reconstruction method~\citep[][see Figure 2 bottom left]{Horowitz19}.

CLAMATO utilizes star-forming galaxies and quasars as background sources to probe the Lyman-alpha (\Lya) forest to determine the large-scale hydrogen distribution at $z\sim 2.05-2.55$ centered at $10^{h}00^{m}34.21^{s}$, $+02^{\circ}17'53.49''$ (J2000). This corresponds to an extent of $438$\,cMpc/h in redshift and $34$\,cMpc/h ($28$\,cMpc/h) in R.A. (Dec.).
The CLAMATO survey was conducted with the LRIS spectrograph on the Keck-I Telescope, yielding around 320 galaxy and quasar background sources in a $\sim 0.2$\,deg$^2$ patch of the COSMOS field. The Lyman-alpha forest fluctuation is determined for each sightline by dividing out the estimated source continuum and assumed mean Lyman-alpha transmission \citep{Lee12}.

While earlier Lyman-alpha forest based reconstruction efforts~\citep{Lee14, Lee18, Newman20} used the Wiener filtering technique~\citep[see e.g.][]{Pichon01, Caucci08} for reconstruction of the large-scale density field, cosmosTNG uses the more recent TARDIS method~\citep{Horowitz19, Horowitz21, Horowitz21b} allowing the reconstruction of the \textit{initial} density field from the CLAMATO survey. The Tomographic Absorption Reconstruction and Density Inference Scheme (TARDIS) utilizes maximum likelihood methods to reconstruct the underlying initial density field with a fast nonlinear gravitational model. A forward model computes the mock \Lya forest measurements and galaxy populations by evolving a given initial density field with the fastPM/flowPM code~\citep{Feng16, Modi21} to the target redshift. A similar forward model based reconstruction approach has been studied in~\citet{Porqueres19, Porqueres20} utilizing a Hamiltonian Monte Carlo solver, albeit requiring a strong cosmological prior and larger computational demand.

For the \Lya forest transmission, the Fluctuating Gunn-Peterson Approximation (FGPA) yields the evolved density field. The improved TARDIS-II allows galaxies to be incorporated into the reconstruction, which we do so by including galaxy redshifts taken from zCOSMOS~\citep{Lilly07}, including unpublished zCOSMOS-Deep data (Kashino et al., in prep). The galaxy overdensity field is modeled using a linear and quadratic bias term in Lagrangian space, effectively following the approach in~\cite{Modi20}.  The log-likelihood for the \Lya forest and galaxy population is taken as a prior term and the sum of the squared differences between the respective observed and model overdensity field in each pixel. The joint log-likelihood is given by their sum. The likelihood is iteratively maximized by running a sequence of forward models using the limited-memory Broyden-Fletcher-Goldfarb-Shanno (LBFGS;~\citet{Liu89}) optimization algorithm\footnote{In the current reconstruction, we find that zCOSMOS galaxies play only a minor role for the resulting initial conditions.}.

Each TARDIS forward step is carried out on a $128^3$ uniform grid with a spatial resolution of $1.0$\,cMpc/h. The resulting reconstructed density field centers on the CLAMATO survey footprint in R.A. and Dec. at a given redshift. In order to cover the whole CLAMATO survey volume across the redshift range, five overlapping, equally-spaced subvolumes are run. These are centered at 59, 147, 235, 323, and 411 cMpc/h in the redshift direction. For each subvolume, we discard the first and last $5$\,cMpc/h. This leaves an overlap of 30\,cMpc/h between each subvolume. We linearly interpolate the density field in the overlapping region by averaging the densities weighted by the distance to the respective subvolume border. We truncate the stitched density field at $450$\,cMpc/h. Overall, the end result is a $128\times 128\times 450$ grid centered on the CLAMATO survey footprint in R.A. and Dec.\footnote{Note that along the redshift direction, only the first $438$\,cMpc/h cover the constrained CLAMATO redshift range $z=2.05-2.55$.}

From the TARDIS maximum a posteriori realization of the density field, we extract a $128^3$ cube for the initial density field of our cosmological simulation. We select a subvolume in order to reduce the total computational cost of the project, although initial conditions for the entire $450$\,cMpc/h volume can be run in the future. The particular subvolume was chosen for two reasons: minimal contamination from overlap, and coverage of the well-studied zFIRE proto-cluster region.

Next, we convert this density field to a high-resolution set of particles for the initialization of the \textsc{AREPO} simulations. Conceptually, this is done by sampling the density field supplemented by random small-scale fluctuations according to the concordance cosmology model, then applying the Zeldovich approximation to obtain the initial particle positions and velocities. Practically, we create a custom version of the MPI-distributed N-GenIC code~\citep{Springel05}. The modified version improves memory efficiency for non-cubical simulation volumes and allows us to fix the large-scale density field and add small-scale fluctuations according to the linear power spectrum.

We start by creating an unperturbed particle distribution, where the target volume is embedded into a larger periodic volume. The periodic volume is cubical with side length $L$ with $L=512$\,Mpc/h. The cubical volume is sampled with $N_{\rm base}^3$ uniformly spaced particles on a Cartesian grid with N$_{\rm grid}=128$, inside of which we embed the $34\times 28\times 118$\,cMpc$^3$/h$^3$ subvolume at our target resolution in the box center. Between the base and target resolution, all intermediate mass resolution levels differ by a factor of eight in mass. We pad the target volume with at least two boundary particles per intermediate resolution level between target and base resolution.

We then upsample the previously reconstructed density field using linear interpolation to a uniform grid with $N_{\rm grid}=2\cdot N_{\rm base}$ elements in each dimension covering the periodic volume $L^3$. We compute the Fourier transform of this field to filter out any unconstrained small-scale power above a cutoff frequency k$_{\rm cutoff}=1$\,h/cMpc. Particle displacements and velocities are then derived using the Zeldovich approximation~\citep{Zeldovich70} at the unperturbed particle positions. The displacements and velocity shifts are applied to all particles within the $128^3$\,cMpc$^3$/h$^3$ input data cube. As such, we also capture large-scale modes outside of the CLAMATO field as reconstructed by TARDIS. Particles outside of the loaded density field do not receive any displacement or velocity. 

Next, we add small-scale density fluctuations above the cutoff frequency. A uniform complex grid representing the Fourier transform of this density field is initialized to zero. This grid has an extent of $D_{x,y,z}=(64, 64, 256)$\,cMpc/h covering the target volume with a grid size of N$_{x,y,z}=2^{1+\mathrm{ceil}\left(\log_2(D_{x,y,z}/\Delta)\right)}$. For all modes with $k>k_{\rm cutoff}$, we draw random numbers from a Gaussian distribution under Hermitian constraints to generate the Fourier transform of a real Gaussian random field~\citep{Sirko05}. Properly normalized, we multiply this field with the square root of the linear power spectrum to obtain a realization for the small-scale fluctuations. The linear matter transfer function is computed using the CAMB code~\citep{Lewis00} assuming Planck15 cosmological parameters (see Section~\ref{sec:conventions}). Displacements and velocities are added for these particles at the target resolution. To match the mass target resolution, we use a sampling factor $0.5 < f < 1$ ($\Delta_\mathrm{sampling}=\Delta/f$) and a Cloud-in-Cell interpolator at the sampled positions.

The resulting initial condition at $z=127$ contains total matter particles. Upon initialization of the simulation, each original particle is split into a dark matter particle and gas cell and displaced from one another by half the mean spacing, while conserving the center of mass of the pair.

Figure~\ref{fig:power_constrained} shows the power spectrum within the cosmosTNG simulation volume. The blue line shows the power spectrum of the TARDIS reconstruction, while the CAMB-based linear expectation is shown in orange. The TARDIS reconstruction power drops off around the chosen cutoff frequency $k\sim 1$\,h/cMpc, reflecting the limited small-scale constraints as well as the reconstruction grid resolution. The chosen cutoff frequency coincides well with the constrained power dropping below the linear expectation. The colored shaded areas reflect the regimes within which we take the TARDIS reconstruction and random realization respectively.

\begin{figure*}
    \includegraphics[width=\textwidth]{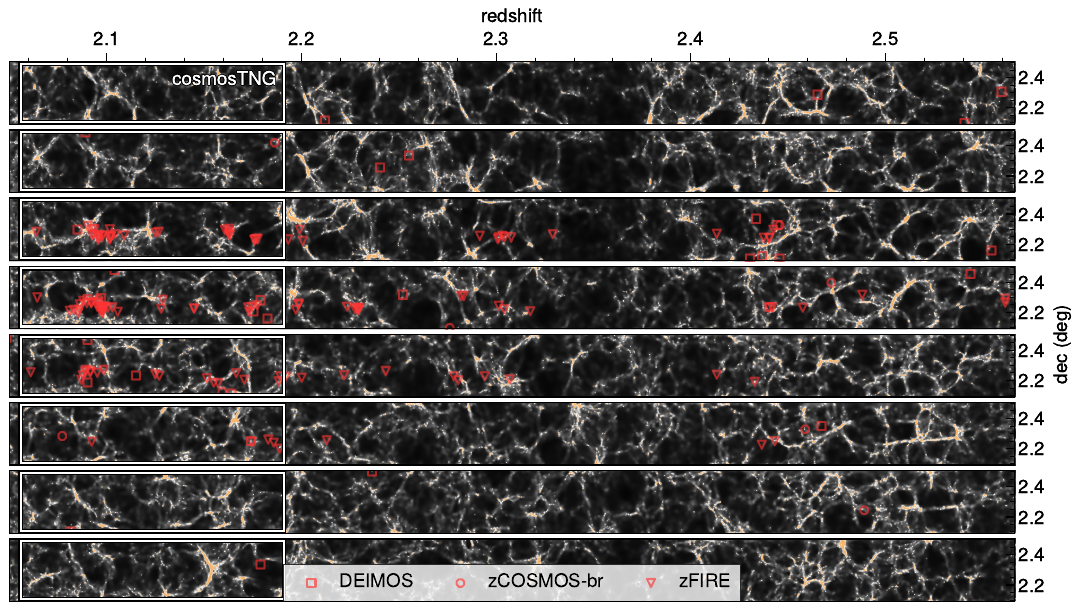}
    \includegraphics[width=\textwidth]{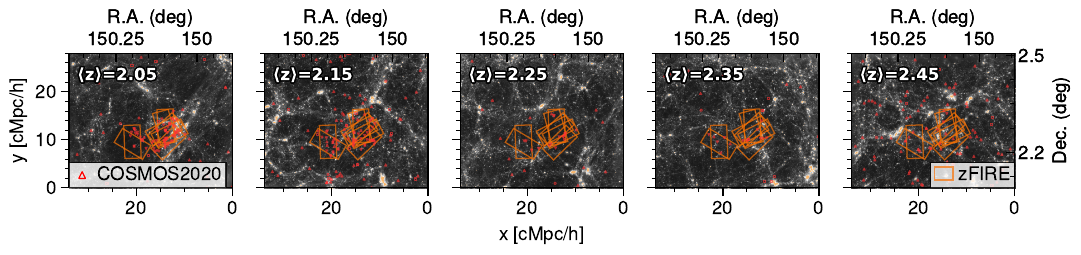}
    \caption{Visualization of the full constrained region evolved to $z\sim 2$ from which cosmosTNG is drawn. \textbf{Top:} Eight consecutive projections of 4\,cMpc/h thick projection of the dark matter density field showing the redshift (declination) on the x-axis (y-axis). \textbf{Bottom:} Five projections of the same dark matter density field as a function of right ascension and declination at increasing redshifts. Colored symbols show observed galaxies from spectroscopic redshift surveys. In orange rectangles, we show the zFIRE survey footprint. In the lower panel, we additionally show photometric detections from the COSMOS2020 catalog as hexagons.}
    \label{fig:cosmosTNG450}
\end{figure*}

Using the previous output, we are ready to perform the evolution step using the \textsc{AREPO} code. In the following, we show and discuss a dark matter only run of the full CLAMATO volume, before introducing the TNG model (Section~\ref{sec:tngmodel}) for the hydrodynamical runs. The dark matter only evolution of the full CLAMATO volume using \textsc{AREPO} is shown in Figure~\ref{fig:cosmosTNG450}, where we visualize the evolved initial conditions at redshift $z\sim 2.3$. In the top panel, we show eight consecutive 4\,cMpc/h thick projections stepping through the right ascension direction. We additionally show observed galaxies from the spectroscopic DEIMOS~\citep{Hasinger18}, zCOSMOS~\citep{Lilly07} and zFIRE~\citep{Nanayakkara16} surveys. The cosmosTNG simulation subvolume is shown as the rectangular region spanning from $z\sim 2.06$ to $z\sim 2.19$, containing the prominent zFIRE protocluster \citep{Spitler12,Nanayakkara16}.
In the bottom panel, we show five $\Delta z\sim 0.1$ projections of the dark matter density field with redshift along the line-of-sight direction. Orange rectangles show the non-trivial footprint of the zFIRE survey -- the largest source of spectroscopic galaxies in this region. We also mark photometric detections from the COSMOS2020 catalog \citep{Weaver22} as hexagons. 

\subsection{The TNG Galaxy Formation Model}
\label{sec:tngmodel}

The TNG galaxy formation model is implemented within the \textsc{AREPO} code~\citep{Springel10}, which solves the coupled equations for self-gravity and ideal, continuum magnetohydrodynamics~\citep{Pakmor11}. The physical domain is discretized with a moving Voronoi tessellation, which allows for a quasi-Lagrangian description of the gas dynamics. The TNG model has been used in several large-volume projects across different regimes and redshifts: the TNG100 and TNG300 simulations of IllustrisTNG \citep{Springel18,Nelson18,Pillepich18,Naiman18,Marinacci18}, the TNG50 volume \citep{Pillepich19,Nelson19a}, THESAN \citep{Kannan22,Garaldi22,Smith22a} MillenniumTNG \citep{Pakmor23}, and TNG-Cluster \citep{Nelson23b}. We briefly summarize the TNG model physics here \citep[see][]{Weinberger17,Pillepich18a}.

The TNG model includes the processes essential for galaxy formation: primordial and metal-line cooling, heating and ionization from an ultraviolet background (UVB;~\citet{Faucher-Giguere09}, FG11 version), star formation above a density threshold~\citep{Springel03}, stellar feedback driven galactic winds, stellar population evolution and chemical enrichment from supernovae Ia, II and AGB stars \citep{Pillepich18a}, and the seeding, merging, accretion growth, and feedback of supermassive black holes \citep[SMBHs;][]{Weinberger17}.

Stellar feedback combines (i) a treatment of small-scale pressure arising from unresolved supernovae generating the hot-phase of the ISM, together with (ii) a decoupled kinetic wind method that produces galactic-scale outflows~\citep{Springel03}. At injection, the wind velocity is set to be proportional to the local dark matter velocity dispersion at a minimum of $350$\,km/s and a mass loading of $\eta=\dot{M}_{\rm wind} / \dot{M}_{\rm SFR}=2(1-\tau_{\rm wind})e_{\rm wind}/v_{\rm wind}^2$, where $\tau_{\rm wind}=0.1$ is the thermal energy fraction and $e_{\rm wind}$ is a metallicity dependent modulation of the canonical energy available per SNII \citep[see][for details]{Pillepich18a}. The resulting feedback-driven outflows collimate along the minor axes of galaxies, with velocities and mass loadings that decrease with distance~\citep{Nelson19a}.

Supermassive black hole feedback follows the description in~\citet{Weinberger17}. The two-state model switches from a kinetic to a thermal mode feedback for high accretion rates relative to the Eddington limit. In addition, radiative feedback is implemented in the form of photoheating and -ionization from a local AGN ionization field impacting up to $3$ times the virial radius of the host halo~\citep{Vogelsberger14}. The AGN luminosity is proportional to the accretion rate above a target accretion threshold, modified by an obscuration factor~\citep{Hopkins07}. The radiation field is added under the assumption of optically thin gas, scaling as $1/r^2$ from its source, and is subject to the same self-shielding description as for the UV background.

\subsection{The cosmosTNG Suite}

{\renewcommand{\arraystretch}{1.3}
\begin{table}
    \centering
    \begin{tabular}{lccc}
        \hline\hline
        Name & $N_\mathrm{DM}$ & $m_\mathrm{DM}$ & $m_\mathrm{gas}$ \\
        \hline\hline
        cosmosTNG-1 & $\sim 1200^3$ & $7.5\times 10^6$\,M$_\odot$ & $1.4\times 10^6$\,M$_\odot$ \\
        cosmosTNG-2 & $\sim 600^3$ & $6.0\times 10^7$\,M$_\odot$ & $1.1\times 10^7$\,M$_\odot$ \\
        cosmosTNG-3 & $\sim 300^3$ & $4.8\times 10^8$\,M$_\odot$ & $8.9\times 10^7$\,M$_\odot$ \\
        \hline
        \textit{TNG100-1} & $1820^3$ & $7.5\times 10^6$\,M$_\odot$ & $1.4\times 10^6$\,M$_\odot$ \\
        \hline
    \end{tabular}
    \caption{Resolution levels of the cosmosTNG suite compared to the TNG100-1 simulation. TNG100-1 and cosmosTNG-1 have matching mass resolutions. $N_\mathrm{DM}$ is the number of dark matter particles, $m_\mathrm{DM}$ the dark matter particle mass, and $m_\mathrm{gas}$ the average gas element mass. cosmosTNG has a volume of $1.1 \times 10^{5}$\,cMpc$^{3}$/h$^{3}$ compared to a $\sim 4 \times$ larger volume of $4.2\times 10^{5}$\,cMpc$^{3}$/h$^{3}$ in TNG100.}
    \label{tab:resolution}
\end{table}
}

{\renewcommand{\arraystretch}{1.3}
\begin{table}
    \centering
    \begin{tabular}{lccc}
        \hline\hline
        Suffix & N$_\mathrm{gal}$ & SFRD$_\mathrm{z=2}$ & M$_\mathrm{ZFIRE,max}$ \\
        & ($M_\star > 10^8 \rm{M}_\odot$) & [$10^{-2}$\,M$_\odot$/yr/cMpc$^3$] & [$10^{13}$\,M$_\odot$] \\
        \hline\hline
        A & 432 & 8.13 & 2.50 \\
        B & 483 & 9.13 & 2.87 \\
        C & 491 & 8.41 & 1.56 \\
        D & 453 & 8.67 & 2.30 \\
        E & 473 & 8.88 & 2.48 \\
        F & 451 & 8.45 & 1.76 \\
        G & 472 & 8.95 & 2.85 \\
        H & 482 & 8.49 & 2.87 \\
        \hline
          & $467\pm 19$ & $8.64\pm 0.31$ & $ 2.40\pm 0.47$ \\
        \hline
    \end{tabular}
    \caption{List of the eight variations, and a few properties from each volume to highlight the diversity that results from varying the small-scale (unconstrained) power. These are: the number of galaxies with stellar mass above $10^8$\,M$_\odot$ ($N_{\rm gal}$), the average instantaneous star formation rate density (SFRD), and the mass of the most massive halo in the zoom-in region (M$_\mathrm{ZFIRE,max}$; upper right panel of Figure~\ref{fig:vis_overview}). All simulations are run to $z=2$, where these properties are measured. The last line shows the mean and standard deviation of the variation runs.}
    \label{tab:variations}
\end{table}
}

The numerical resolution of the cosmosTNG simulations are given in Table~\ref{tab:resolution}. The primary runs have a dark matter mass resolution of $m_\mathrm{DM}=7.5\times 10^6$\,M$_\odot$ and a gas mass resolution of $m_\mathrm{gas}\sim1.4\times 10^6$\,M$_\odot$ in the high-resolution $34\times 28\times 118$\,cMpc$^3$/h$^3$ subvolume. To study numerical convergence, we also run lower resolution versions of every volume, labeled \mbox{`-2'} (8 times lower mass resolution) and \mbox{`-3'} (64 times lower mass resolution). These resolution levels are equal to those of TNG100, and cosmosTNG-1 has the same resolution as TNG100-1. For each run we save 34 snapshots down to redshift $z=2$.

We run eight variations of cosmosTNG\@. Each has different density fluctuations for the unconstrained, small-scale modes. The variations are denoted by a letter from A to H, and each letter represents a different random seed, used to initialize and draw a pseudorandom number sequence used to set the amplitudes and phases above the cutoff frequency $k_\mathrm{cutoff}=1$\,h/cMpc, see Figure~\ref{fig:power_constrained}. These variations allow us to study the impact of small-scale variations on the galaxy properties while keeping the large-scale structure fixed. A summary of properties for the variations is shown in Table~\ref{tab:variations}.

\subsection{Conventions}
\label{sec:conventions}

We use `p' and `c' to explicitly note comoving or physical coordinates across the paper. In the analysis, unless stated otherwise, stellar masses and star-formation rates are computed in an aperture with 2 arcsecond diameter. Halo masses are taken as the total mass of a Friends-of-Friends group enclosed in a sphere whose mean density is 200 times the critical density of the universe at the respective redshift. For calculating all statistics, we ignore the outer $2\,$cMpc/h of the target volume to mitigate numerical effects in the proximity of low-resolution particles.

For TARDIS and \textsc{AREPO}, we assume a concordance flat $\Lambda$CDM cosmology with $\Omega_{\Lambda,0}=0.6911$, $\Omega_{m,0}=0.3089$, $\Omega_{b,0}=0.0486$, $\sigma_8=0.8159$, $n_s=0.9667$ and $h=0.6774$, consistent with \citet{PlanckCollaboration16}. For mapping from comoving Cartesian coordinates to celestial coordinates and redshift we use a fixed relation $\chi=3874.867$\,cMpc/h and $\mathrm{d}\chi/\mathrm{d}z=871.627$\,cMpc/h at $\langle z\rangle=2.30$ for simplicity.\footnote{This is a minor difference in the cosmology used for FlowPM and \textsc{AREPO} compared to these conversion factors, in order to be consistent with the CLAMATO data release.}

\section{Results}
\label{sec:results}

\begin{figure*}
    \centering
    \includegraphics[width=1.0\textwidth]{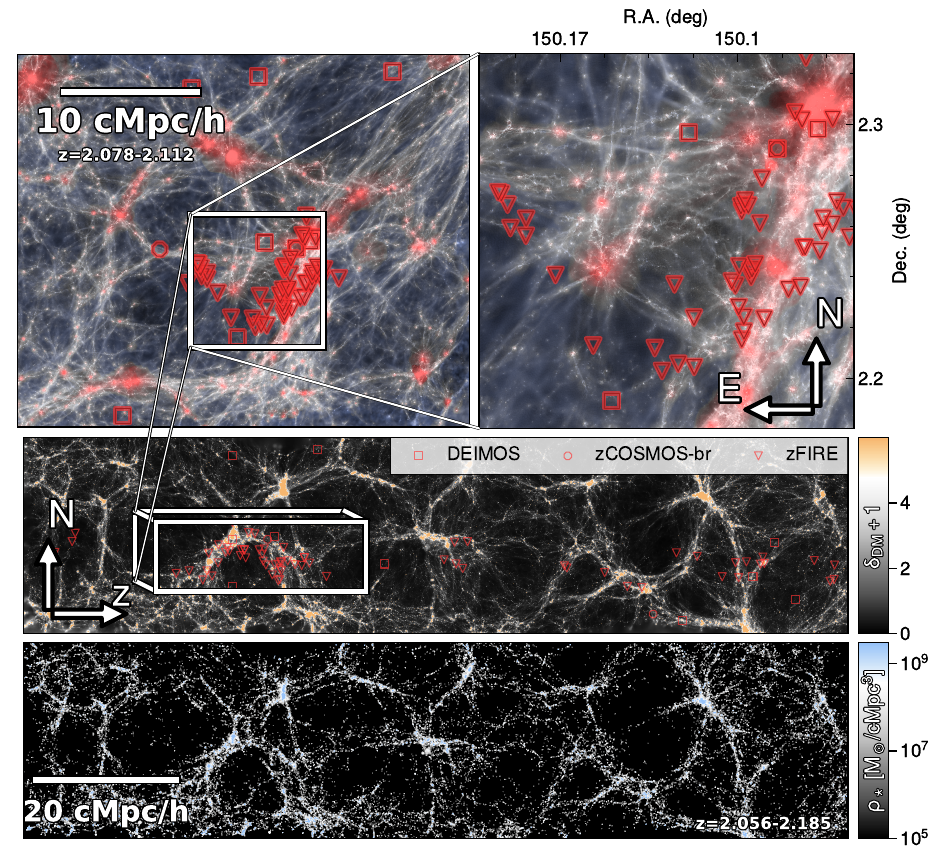}
    \caption{Visual overview of the cosmosTNG-A simulation at $z=2$.
    \textbf{Top panel:} Two 30\,cMpc/h projections along the redshift direction with the right ascension (declination) on the x-axis (y-axis). On the left, we show the CLAMATO footprint, and on the right a zoom-in of a highly overdense region studied by the zFIRE survey. Both projections show a two-dimensional colormap, where blue (red) indicates cold (hot) gas, and white (black) a high (low) density. Red symbols in the upper panels show observed galaxies from various spectroscopic surveys.
    \textbf{Bottom panels:} Dark matter and stellar density projections for a 10\,cMpc/h thick slice through the simulation volume with the redshift (declination) on the x-axis (y-axis). 
    Visually, we find that observed galaxies spatially correlate with the simulated density field down to the reconstruction scale of $~$1\,cMpc/h. This correlation is particularly striking in the redshift direction within the zFIRE region, reproducing a characteristic `arc' shape of the galaxy distribution.}
    \label{fig:vis_overview}
\end{figure*}

In Figure~\ref{fig:vis_overview} we show an overview plot of the simulated cosmosTNG volume that visualizes several different physical properties. The two panels at the top show the density-temperature map of a zoom-in region centered on the zFIRE protocluster region at $z=2$ in the right-ascension-declination plane. Blue (red) colors indicate cold (hot) gas, while white (black) indicates a high (low) density. We see a web of filamentary structures at intermediate temperatures, with hot gas present and abundant at its nodes. In the two lower panels, we show the average dark matter and stellar mass density in the redshift-declination plane.

For qualitative comparison, red markers indicate observed spectroscopic galaxies (top three panels). We generally see good visual agreement between simulated overdensities and observed galaxies on $\sim$\,Mpc scales, near the constrained modes of the initial conditions. Due to irregular survey footprints, the correctness of observed galaxy overdensities is hard to assess in R.A.-Dec. projections. For the redshift-declination projection of the dark matter density however, we see good agreement between observed and simulated structures as visually confirmed by the arc-like structure of the protocluster region. 

\begin{figure*}
    \centering
    \includegraphics[width=1.0\textwidth]{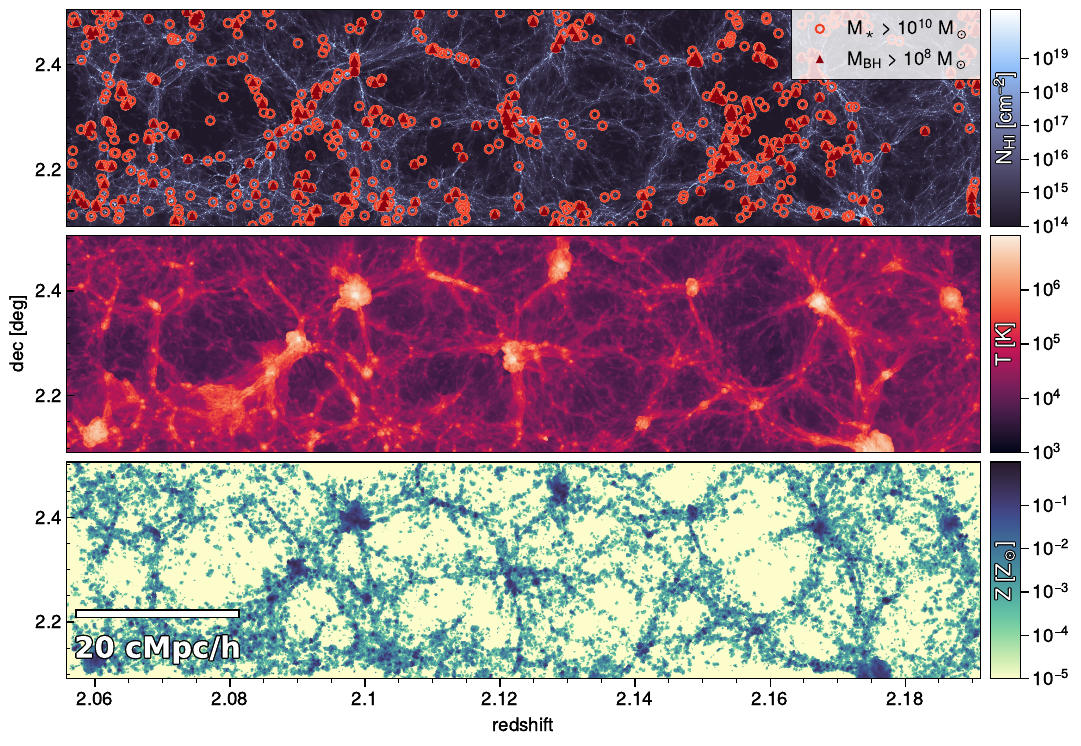}
    \caption{Projections of the neutral hydrogen column density (\textbf{top}), temperature (\textbf{middle}) and metallicity (\textbf{bottom}) along the R.A. direction through the cosmosTNG-A simulation at $z=2$ for a 10\,cMpc/h depth with the redshift (declination) on the x-axis (y-axis). In the top panel, we also mark simulated galaxies and black holes with masses above $10^{10}$\,M$_\odot$ and $10^8$\,M$_\odot$, respectively. Thin filaments of neutral hydrogen span the volume and at its nodes, massive galaxies reside whose AGN heat and metal-enrich their surroundings. The white box in the top panel indicates the zFIRE cluster zoom-in region as in Figure~\ref{fig:vis_overview}.}
    \label{fig:vis_overview2}
\end{figure*}

In Figure~\ref{fig:vis_overview2}, we show projections for the redshift-Dec. plane for the neutral hydrogen column density, gas temperature and gas metallicity. In the top panel, we also show simulated galaxies and black holes with masses above $10^{10}$\msun and $10^8$\msun, respectively. The cosmic web is clearly visible in the various panels, and filamentary structures are particularly pronounced in the neutral hydrogen density maps, covering a large dynamical range: from $<10^{14}$\,cm$^{-2}$ in voids, to filaments often in excess of $10^{17}$cm$^{-2}$ and reaching $>10^{20}$cm$^{-2}$ in large overdensities. The temperature map, as well as galaxy and black hole positions also trace the large-scale structure, albeit more biased towards overdense nodes of the cosmic web, often exceeding $10^{6.5}$\,K in the proximity of massive galaxies. Heated regions near cosmic web nodes are accompanied by metal-enriched gas ($>10^{-1}$\,Z$_\odot$) up to scales of multiple cMpc, indicating AGN-driven outflows out to these scales and beyond.

\begin{figure}
    \includegraphics[width=\columnwidth]{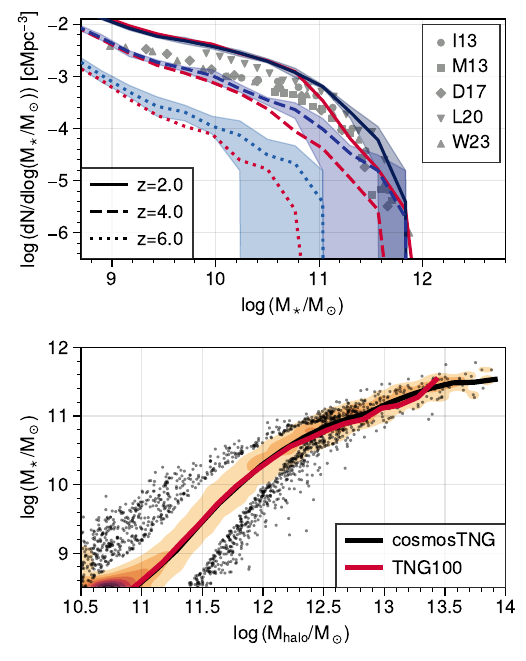}
    \caption{\textbf{Top:} The galaxy stellar mass function from cosmosTNG at redshifts $z=2,4,6$. The semi-transparent bands span the outcomes of different variation runs with the embedded line showing the mean. In red, we show the TNG100 simulation outcomes. We compare to several observational datasets~\citep{Ilbert13, Muzzin13, Davidzon17, Leja20, Weaver23} at face value. These observations span $z\sim 2.0-2.5$ and are shown with gray markers. In comparison to TNG100, cosmosTNG produces more high mass galaxies above $10^{11}$\msun by up to $0.5\,$ dex. This reflects the overabundance of massive dark matter halos in this region (see text). \textbf{Bottom:} The stellar mass to halo mass relation for stellar masses within previous aperture of all central galaxies and the mass of their hosting halo at $z=2$. Orange contours indicate the distribution of stellar masses at fixed halo mass across all cosmosTNG variation runs, while individual points show galaxy outliers. The solid black (red) line shows the median stellar mass for cosmosTNG (TNG100).}
    \label{fig:SMF}
\end{figure}

\subsection{Stellar Mass Function and Stellar-to-Halo-Mass Relation}

We begin our quantitative analysis of cosmosTNG with integral properties of the galaxy population. In the top panel of Figure~\ref{fig:SMF}, we show the galaxy stellar mass function (SMF) at redshifts $z\sim 2,4,6$. We measure the stellar mass within a $2.0$ arcsecond diameter aperture, and including all subhalo stars. The bands show the range of outcomes of the eight different variation runs, and the red lines show the TNG100 result. 
Gray symbols show observational data~\citep{Ilbert13, Muzzin13, Davidzon17, Leja20, Weaver23} for $z\sim 2.0-2.5$.

When comparing to observations, here and through the results section, we use various recent observational datasets to give a sense of the variation due to different observational methods and target selection functions, allowing better face-value comparisons against cosmosTNG and TNG100.

The TNG100-1 result is within $0.5\,$dex of the data at these redshifts, always in the direction of having too many galaxies at a given stellar mass.  We stress that differences in methodology and aperture corrections make quantitative comparisons to data difficult \citep[see also][for previous comparisons of the TNG100 and TNG300 SMFs with data]{Pillepich18}. While generally on the upper end, cosmosTNG and the TNG model are broadly consistent with observational results from~\citet{Leja20}, who employ a continuity model accounting for the redshift evolution of the mass function within the fitting procedure. The level of disagreement between the observational inferences suggests that systematic uncertainties are similar to the level of (dis)agreement seen between the observations and simulations.

Of primary interest, cosmosTNG is consistent with TNG100 only at stellar masses below $10^{11}$\msun. At higher masses, cosmosTNG exceeds the number of galaxies compared to TNG100, as well as general i.e. non-COSMOS field observations. At this high-mass end, the galaxy count in cosmosTNG exceeds that of TNG100 due to a shallower slope of the SMF. This discrepancy is largest at high redshifts and decreases towards $z\sim 2$. This behavior is consistent across all variation runs. Given the identical galaxy formation model and resolution, the constrained initial conditions produce this difference. Specifically, a large abundance of high-mass galaxies is driven by an excess of large-scale density fluctuations (see Figure~\ref{fig:power_constrained}) at $k<1$\,cMpc/h. 
In Figure~\ref{fig:resolutionstudy} of the Appendix we show the halo mass function at $z=2$, and its evolution from the initial conditions. An excess abundance of massive dark matter halos with respect to a random region of the universe directly leads to the $z \sim 2$ excess in the high-mass end of the SMF.

In the bottom panel of Figure~\ref{fig:SMF}, we show the stellar mass to halo mass relation $M_\star-M_\mathrm{halo}$ at $z=2$ for cosmosTNG and TNG100. The relation follows a step power-law scaling up to approximately $M_\mathrm{halo}=10^{12}$\msun after which the relation significantly flattens in line with the SMF drop off. While next to identical at lower masses, there is a mild $0.1-0.2$\,dex higher median relation for cosmosTNG compared to TNG100 above $M_\mathrm{halo}=10^{12}$\msun, contributing to the massive galaxy overabundance in the SMF.

\begin{figure}
    \includegraphics[width=\columnwidth]{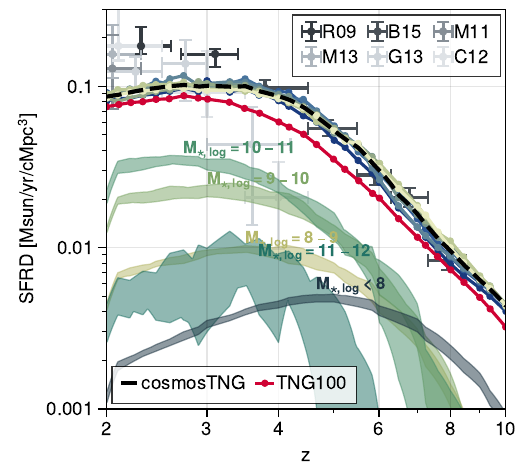}
    \caption{The cosmic star formation rate density (SFRD) in cosmosTNG between z=$2$-$10$. The colored lines show the individual realizations, and the black line shows the average. In red we show the TNG100 simulation for comparison. Semi-transparent bands show the range contributions of different stellar mass ranges to the overall SFRD budget across different variation runs. Gray points indicate observational data from various surveys~\citep[abbrv. R09, M11, C12, M13, G13, B15 for][]{Reddy09,Magnelli11, Cucciati12, Magnelli13, Gruppioni13, Bouwens15}, with some data from~\citet{Madau14, Bouwens15}. We compare this data to cosmosTNG at face value, i.e. without replicating the procedure for the observational inferences. The SFRD of cosmosTNG is overall higher, and peaks at earlier redshift, than the (field) result of TNG100, suggesting that $z\sim 2.1$ galaxies within this region of COSMOS have undergone a particularly active assembly history.}
    \label{fig:SFRD}
\end{figure}

\subsection{Star Formation Rate History}

In Figure~\ref{fig:SFRD}, we show the cosmic star-formation rate density (SFRD) as a function of redshift. The SFRD is calculated as the sum of instantaneous star-formation in each gas cell divided by the simulated volume at target resolution. The black dashed line shows the mean of all 8 variation runs and individual colored lines show the evolution for individual runs. The red line shows the same line for the TNG100 simulation (see \citet{Shen22} for dust modeling and obscured star formation effects in the SFRD, and \citet{Pillepich18a} for the impact of physical model choices on the SFRD, in the context of the TNG simulations).

The eight variation runs deviate by less than 20-30\% from one another, with relative differences being the largest around $z\sim 7$. Strikingly, the cosmosTNG mean and all variation runs are consistently higher than TNG100 at all redshifts. The relative difference between cosmosTNG and TNG100 peaks around $z\sim 4$ with cosmosTNG having nearly twice as much star-formation taking place. Furthermore, cosmosTNG reaches its maximum SFRD earlier at $z\sim 3.5$ compared to $z\sim 3$ for TNG100.

Gray markers show observationally inferred SFRD values \citep[][and others, see Figure caption for details]{Madau14}. The SFRD inferred, particularly at cosmic noon between $z\sim 2-3$, shows significant variation across observational studies. At face value, these are consistent with cosmosTNG down to its maximum at redshift $z\sim 3.5$. In contrast to the TNG model, many observationally inferred SFRDs keep rising towards $z\sim 2$, leading to discrepancy of up to roughly a factor of two depending on the dataset at this redshift \citep[however, see][for systematics and further discussion, including a relatively flat inferred SFRD from $z\sim 2$ to $z\sim 3.5$]{Enia22}.

In Figure~\ref{fig:SFRD} we also show how different halo masses contribute to the global SFRD \citep[following][with Illustris]{Genel14}. We sum all star-formation associated with subhalos in a given stellar mass range, and plot the range of outcomes between different variation runs as shaded bands. Here, stellar masses are computed summing over all stars belonging to a subhalo.

We find that galaxies with stellar masses from $10^{10}$ to $10^{11}$\msun dominate between $2<z<7$, even though contributions for galaxies from $10^9$ to $10^{10}$\msun have nearly equal weight. Interestingly, the overall turnover at $z\sim 3.5$ does not occur for these mass bins, but is instead driven by galaxies with lower as well as higher mass. Star formation in halos between $10^{8}$ to $10^{9}$\msun and $10^{10}$ to $10^{11}$\msun equally contribute roughly $10\%$ each to the overall budget at $z\sim 3.5$, where they both peak. However, lower mass halos already contribute more substantially at higher redshifts. The SFRD of low mass halos $< 10^8$\msun peaks at $z \sim 5$, and declines towards higher as well as lower redshift.

\subsection{Galaxy Size-Mass Relation}

\begin{figure}
    \includegraphics[width=\columnwidth]{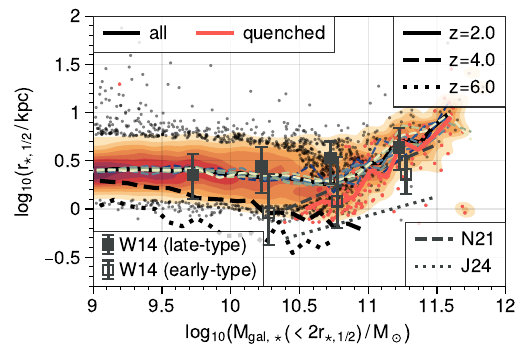}
    \caption{The size-mass relation in cosmosTNG at $z=2$, combining all central galaxies regardless of type. Orange contours indicate the distribution of sizes at fixed stellar mass across all cosmosTNG variation runs, while individual black points show outliers. The black line shows the median size at fixed stellar mass. The dashed (dotted) line shows the median at redshift $z=4$ ($z=6$), while thin colored dashed lines show the individual variation runs.
    Finally, red dots and the red line show quenched galaxies only (see text).
    We define galaxy size as twice the stellar half-mass radius, and here measure stellar masses summing all stellar populations within this radius.
    The observational data span $z=2.0$ to $z=2.5$ \citep{vanderWel14}, where open symbols show early-type galaxies and filled symbols show late-type galaxies. Additionally, we show the linear scaling relations for quenched galaxies derived from observations~\citet{Nedkova21} and~\citet{Ji24} with gray lines. The observational data is compared as-is, without any further observational mock post-processing of cosmosTNG.}
    \label{fig:SMR}
\end{figure}

In Figure~\ref{fig:SMR}, we show the size-mass relation for central galaxies in cosmosTNG at $z=2$. Orange contours show the distribution of galaxy sizes at fixed galaxy mass. Dots represent individual galaxies that are outliers with respect to the main population. Here we measure galaxy size as twice the stellar half mass radius, and galaxy mass as the stellar mass contained within a sphere of this size, in order to be self-consistent. The solid black line shows the median relation, while colored dashed lines show the $z=2$ median for individual variation runs. 

For low mass galaxies, the distribution and median of galaxy sizes is roughly constant with a median size of $\sim 2.5$\,pkpc \citep[resolution convergence at the low-mass end and smaller galaxy sizes in TNG50 are discussed in][]{Pillepich19}. At a stellar mass of $10^{10.8}$\msun sizes start to increase rapidly, towards a power-law slope of $\sim 2/3$. The turning point at $10.8\,$\msun coincides well with the turnover for the sSFR(M$_\star$) relation in Figure~\ref{fig:ssfr_vs_mstar}.

Red points and the red line show quenched galaxies and their median relation only, visible for $M_\star > 10^{10.5}$\msun. We see that quenched galaxies in cosmosTNG, as in observations, are smaller than the overall sample by $0.2$-$0.4$\,dex, with the largest difference at M$_\odot\sim 10^{10.5}$\msun. The difference vanishes towards larger masses as quenched galaxies start to dominate the overall sample \citep[see][for a discussion on the broad agreement of TNG galaxy sizes with data, as a function of redshift as well as split by star-formation activity, representing an important model validation outside of the scope of calibrated observables]{Genel18}.

The dashed (dotted) black lines show the median galaxy size at higher redshift, $z \sim 4$ ($z \sim 6$). At earlier times, galaxies at fixed mass are smaller, and there is a pronounced trend of smaller size with increasing mass \citep[see also][]{Constantin23, Karmakar23, Du24}.

We furthermore show observational data from~\citet{vanderWel14} for various stellar mass bins, for the late-type galaxy population (filled symbols), and early-type galaxies. Gray lines show the linear scaling relations for quenched galaxies as inferred from observations at $z=1.5-2.0$ and $z=2.0-2.5$ in~\citet{Nedkova21} and~\citet{Ji24} respectively. While we generally find good agreement with~\citet{vanderWel14}, the largest discrepancy exists at stellar masses close to the turning point of $10.8$\msun. The observations indicate a more constant slope of size growth with mass across this range. Finally, we see sizes at the high-mass end in good agreement with~\citet{Nedkova21}, while the near-infrared observations from~\citet{Ji24} generally point towards substantially smaller galaxy sizes and a shallower trend with mass.

\subsection{Specific Star Formation Rates, Quenching and Gas Budget}

\begin{figure*}
    \includegraphics[width=\columnwidth]{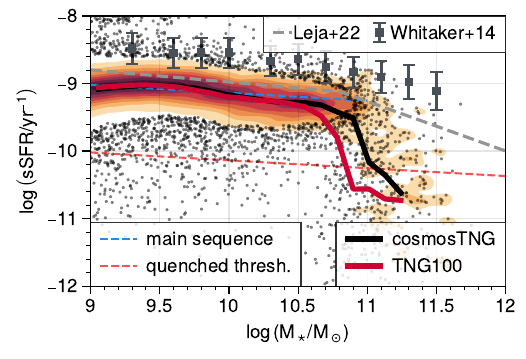}
    \includegraphics[width=\columnwidth]{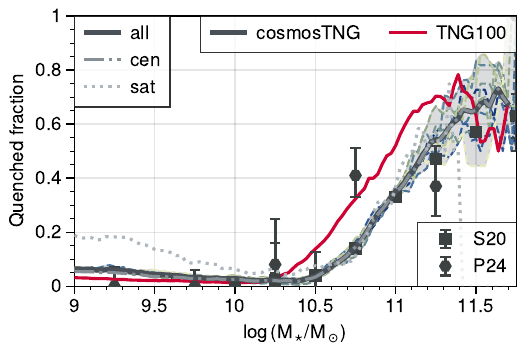}
    \caption{Star formation activity of galaxies. \textbf{Left:} The specific star formation rate (sSFR) as a function of stellar mass for central galaxies in cosmosTNG at $z=2$. Annotations and markers as in Figure~\ref{fig:SMF} bottom panel. Measurements are within 2 arcsecond diameter apertures and with a $100$\,Myr averaged SFR. Gray markers show inferred sSFR from observations in~\citet{Whitaker14}. The dashed gray line shows the sSFR ridge from observational modeling in~\citet{Leja22}. \textbf{Right:} The quenched fraction as a function of stellar mass in cosmosTNG and TNG100 at $z=2$. Errorbars show observations from~\citet{Sherman20a,Park24} around cosmic noon. The observations are compared as-is, i.e.\ without reproducing the methodology as used in the simulation.
    Quenched galaxies are defined by having a star-formation rate by at least 1 dex below the main sequence (see text). The colored lines show the individual realizations, and the dark solid line shows the average for all galaxies. The dashed (dotted) line shows the average for central (satellite) galaxies. A sizable population of quenched (central) galaxies with M$_\star$ > $10^{10.5}$\msun is already present by $z=2$. We see a lower quenching for cosmosTNG galaxies compared to TNG100.}
    \label{fig:ssfr_vs_mstar}
\end{figure*}

In Figure~\ref{fig:ssfr_vs_mstar}, we show the star formation activity of galaxies in terms of the specific star formation rate (sSFR; left panel) and the quenched fraction (right panel) as a function of stellar mass in cosmosTNG at $z\sim 2$. The sSFR and stellar masses are measured in a 2 arcsecond diameter aperture, and the sSFR is averaged over the last 100\,Myr prior to $z=2$.

We determine the quenched fraction by counting those galaxies with a star-formation rate that is $1$\,dex below the main sequence (SFMS). The simulated main sequence is determined with an iterative scheme following~\citet{Donnari21}, where we compute the median star-formation rate in stellar mass bins of 0.2\,dex, excluding all galaxies 1\,dex below this median value as quenched. This procedure is then repeated until convergence. We then take all non-quenched galaxies in the stellar mass range from $10^9$ to $10^{10.2}$\msun, fit a linear relation in the $\log{\mathrm{M}_\star}-\log{\mathrm{SFR}}$-plane, and label all galaxies 1\,dex below this line as quenched. For reference, the blue dashed line shows the SFMS, and the red dashed line shows the threshold for quenched galaxies.

For the left panel of~\ref{fig:ssfr_vs_mstar}, we include central galaxies from all 8 variation runs. Red contours show the sSFR distribution at fixed stellar mass, while black points show individual galaxies where the sampling is sparse. We indicate the median sSFR relation for cosmosTNG (TNG100) with a solid black (red) line. The median lines for cosmosTNG and TNG100 closely align and follow the SFR ridge seen in the contours. 

We overplot the SFR ridge from observational modeling in~\citet{Leja22} with the dashed gray line. Overall, it is $\sim 0.2\,dex$ higher than the sSFR ridge in cosmosTNG. Inferred observational values from~\citet{Whitaker14} are higher by around $0.5$\,dex.
Similar to the parametric drop of the SFR ridge slope at $10^{10.6}$\msun in~\citet{Leja22}, we find a decrease in cosmosTNG around $10^{10.8}$\msun, albeit with a substantially steeper slope. The sSFR distribution becomes significantly broader for these high-mass galaxies \citep[see also][]{Ilbert15}.

In the right panel, we show the quenched fraction of all (central, satellite) galaxies in cosmosTNG at $z\sim 2$ as gray solid (dash-dotted, dotted) line. The thin dashed lines show the quenched fraction for the individual variation runs, and the solid red line the quenched fraction in TNG100 \citep[see][]{Donnari19,Gupta21}. We show observational measurements for the quenched fraction~\citep{Sherman20a,Park24}.\ \citet{Sherman20a} compute the quenched fraction for a sample of roughly $\sim 30,000$ massive galaxies, combining photometric data from DECam with other bands to perform spectral energy distribution fitting in large area surveys covering a total of $17.5$\,deg$^{2}$.\ \citet{Park24} shows a recent JWST result of quenched fractions using deep NIRSpec spectra from the Blue Jay survey, overlapping in parts with cosmosTNG\@. Both observational datasets use the same quenched definition (1 dex below the SFMS) as here. For the observed redshift range ($2<z<2.5$ for~\citet{Sherman20a} and $1.7<z<3.5$~\citet{Park24}), there is a twenty percent point higher quenched fraction at $z\sim 10^{10.7}$\msun in~\citet{Park24}. Alternatively, the difference can be interpreted as a $\sim 0.25$\,dex shift towards lower stellar mass bins at fixed quenched fraction. This highlights potential effects from different target selections and available data for SED fitting, while environmental effects might also be at play. 4 of the 16 quenched galaxies in~\citet{Park24} fall into the zFIRE protocluster covered by cosmosTNG.

cosmosTNG is in good agreement with~\citet{Sherman20a}, while results from~\citet{Park24} suggest an onset of quenching at lower masses, more similar to that of the TNG100 simulation. We note that if the difference between~\citet{Sherman20a} and~\citet{Park24} is a true environmental effect, it goes in the opposite direction as seen in our two simulations.

While the quenched fraction curves for both cosmosTNG and TNG100 follow a very similar shape, there is a significant offset between $M_\star \simeq 10^{10.5}$\msun and $M_\star \simeq 10^{11.5}$\msun. On one hand, such galaxies in cosmosTNG reach the quenched fraction found in TNG100 at a $0.4$\,dex higher stellar mass. Alternatively, galaxies of a given stellar mass have a $\sim 20$ percent lower quenched fraction. We investigate this compelling difference in quenched fractions between cosmosTNG and TNG100 in the discussion.

\begin{figure}
    \includegraphics[width=\columnwidth]{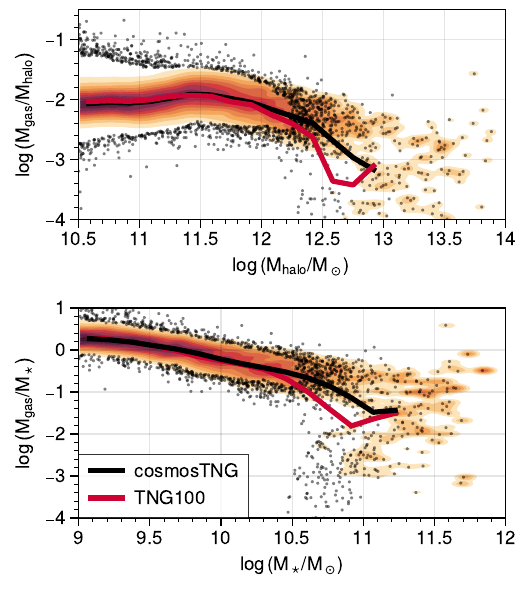}
    \caption{The gas fraction for all central galaxies relative to their hosting halo mass \textbf{(top)} and their stellar mass \textbf{(bottom)} in cosmosTNG at $z=2$. Annotations and markers as in Figure~\ref{fig:SMF} bottom panel. Gas and stellar mass are calculated within twice the stellar half-mass radius. We see an increased drop and scatter for gas fractions above $M_\mathrm{halo}\gtrsim 10^{12}$\msun and $M_\star>10^{10.5}$\msun, in line with changes in the $M_\star$($M_\mathrm{halo})$ and sSFR($M_\star$) relations. Above these mass thresholds, cosmosTNG galaxies show a larger gas fraction on average.}
    \label{fig:gasfrac}
\end{figure}

In Figure~\ref{fig:gasfrac}, we show the gas fraction relative to the central galaxy host halo mass (top) and galaxy stellar mass (bottom). The stellar and gas masses are computed within twice the stellar half-mass radius of the central galaxies. In the upper panel, we see the median gas fraction $M_\mathrm{gas}/M_\mathrm{halo}$ staying roughly constant between $10^{10}$ to $10^{12}$\msun at $\sim 1\%$, while there is a modest enhancement around $10^{11.5}$\msun and increased scatter towards lower stellar masses. Above $10^{12}$\msun the median quickly drops to $\lesssim 0.1\%$ at $10^{13}$\msun with the scattering widening again. The drop at $10^{12}$\msun coincides with the flattened $M_\star$-$M_\mathrm{halo}$ relation (see Figure~\ref{fig:SMF}), while latter relation shows significantly less scatter beyond this mass threshold. The lower panel shows the gas fraction in terms of the stellar mass, showing a monotonously decreasing $M_\mathrm{gas}/M_\star$, dropping below unity around $10^{10}$\msun. At $10^{10.7}$\msun, we see a substantial drop similar to sSFR(M$_\star$) in Figure~\ref{fig:ssfr_vs_mstar}.

\subsection{Galaxy Number Densities}

\begin{figure*}
    \includegraphics[width=\textwidth]{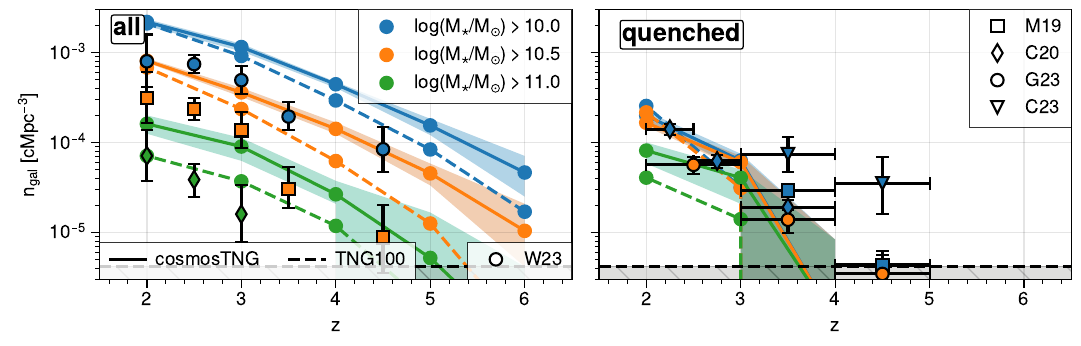}
    \caption{The number density of all (i.e. central and satellite) galaxies as a function of redshifts above a given stellar mass threshold in cosmosTNG-1. Shaded regions show the variation in number density across all random variations. The solid lines show the respective mean. \textbf{Left}: all galaxies. Observational data from~\citet{Weaver23}. \textbf{Right}: quenched galaxies. Observational data for quenched galaxy densities are taken from~\citet{Merlin19, Carnall20, Gould23, Carnall23}. Observations are compared at face value without observational mock post processing of cosmosTNG\@. The hatched region at the bottom shows the number density $1/\mathrm{V}_\mathrm{sim}$, i.e. one galaxy within the cosmosTNG simulation volume $\mathrm{V}_\mathrm{sim}$.}
    \label{fig:ngal_vs_redshift}
\end{figure*}

In Figure~\ref{fig:ngal_vs_redshift}, we show the number density of galaxies as a function of redshift. In the left (right) panel, we show the number density for all (quenched) galaxies. Blue, orange and green lines give the result for three different lower galaxy stellar mass thresholds of $10^{10.0}$, $10^{10.5}$, and $10^{11.0}$\msun. Bands show the variation in the number density across different variation runs. For comparison, dashed lines show the number densities of galaxies from TNG100.

In the left panel, at redshift $z=2$, we find the number density of massive galaxies $\geq 10^{11}$\,\msun in cosmosTNG to exceed those in TNG100 by a factor of 2-3, consistent with the shape of the SMF in Figure~\ref{fig:SMF}. This difference disappears at lower mass thresholds. In contrast, it is generally strong at all mass thresholds for higher redshifts. For quenched galaxies (right panel), differences between cosmosTNG and TNG100 are similarly pronounced for $M_\star \geq 10^{11}$\,\msun for $z=2$, and at all masses for $z \geq 3$.

The existence and abundance of massive galaxies at high redshift, particularly massive quiescent galaxies, has recently been emphasized as a possible source of tension between observations and models within COSMOS at $z\gtrsim 3$~\citep{Forrest24}. We therefore show several observational constraints with markers and their associated error bars. For the number density of all galaxies, we integrate the parameterized double Schechter fit from~\citet{Weaver23}, and show a conservative upper limit derived from the parameter errors of the Schechter fit. For quenched galaxies, we directly take densities and their uncertainties inferred from various observations~\citep{Merlin19, Carnall20, Gould23, Carnall23}. The JWST NIRCam observations presented in~\citet{Carnall23} within the Extended Growth Strip (EGS) show a significant higher estimate than previous observational studies for the number counts of quenched galaxies at $3<z<4$ and particularly $4<z<5$. The authors demonstrate the high counts to be associated with JWST instrument capabilities rather than selection within EGS\@, although further spectroscopic follow-ups might be needed~\citep{Forrest24a}.

We find the observationally inferred number density of galaxies to be broadly consistent at $z=2$ for massive galaxies in TNG model simulations. At lower stellar mass thresholds and higher redshifts, observations are systematically lower compared to cosmosTNG and TNG100. That is, the simulations have more than a sufficient number of massive galaxies at early times, possibly even too many. For example, the number density of galaxies $10^{10.5}$\,\msun from~\citet{vanderWel14} at $z\sim 4.5$ is a factor of $\sim 2$ ($\sim 8$) lower than TNG100 (cosmosTNG).

At the same time, the qualitative trend is the opposite for the quenched galaxy population. At $z\sim 2-3$ the simulations and observational inferences are broadly consistent. In contrast, cosmosTNG and TNG100 yield systematically lower number densities than observed for $z\geq 3.5$. However, at $z > 4$ in particular, the volumes of TNG100 and cosmosTNG are too small to expect such systems (horizontal dashed black line). Number densities derived from narrow redshift ranges in fields as small as COSMOS clearly suffer from cosmic variance at the high-mass end. However, the large number density of~\citet{Carnall23} remains in tension with the TNG model and our definitions: these results echo a debate that has become very urgent and popular in the last years~\citep[e.g.][]{Valentino23}. However, we have not here made any detailed mocking of the tracers, definitions, or methods used to determine quiescence in data, making this a face-value comparison.

\subsection{Galaxy and Black Hole Co-evolution}

\begin{figure}
    \includegraphics[width=\columnwidth]{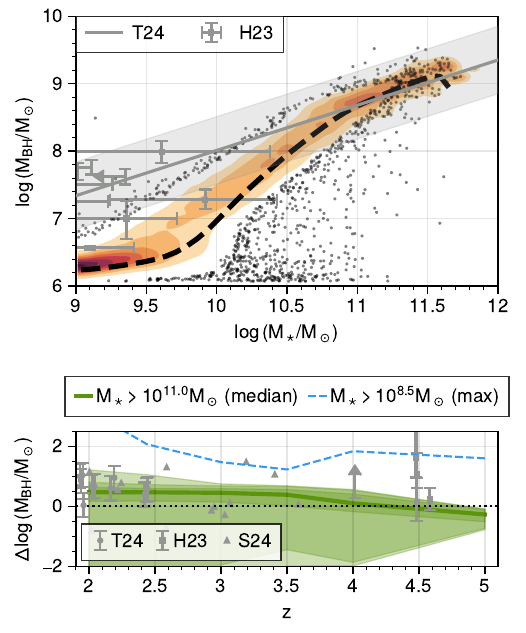}
    \caption{The relation between supermassive black hole mass and galaxy stellar mass, and its evolution with time. \textbf{Top}: The $M_\mathrm{BH}$-$M_*$ relation at $z=2$ in cosmosTNG. The gray line shows the linear relation and the $1-\sigma$ scatter from~\citet{Tanaka24}. Annotations and markers as in Figure~\ref{fig:SMF} bottom panel. \textbf{Bottom}: The evolution of the offset from the local $M_\mathrm{BH}$-$M_*$ relation as described in~\citet{Tanaka24} based on~\citet{Haring04, Bennert11} for the local relation. Points with error bars show high-redshift observations from~\citet{Harikane23} and~\citet{Tanaka24}. Gray triangles show observations from~\citet{Sun24}. All observations are compared as-is, i.e.\ without corrections for the respective observational methodology used when compared to cosmosTNG\@.
    The green line shows the median relation in cosmosTNG for host galaxies with M$_\star>10^{11}$\,M$_\odot$. Shaded regions show the central 68\%, 95\% and all percentiles respectively. The blue dashed line shows host galaxies with M$_\star>10^{8.5}$\,M$_\odot$. The observed galaxies with large black holes masses are rare but possible in cosmosTNG across $z=2-5$.} 
    \label{fig:smbh_gal_evolution}
\end{figure}

In Figure~\ref{fig:smbh_gal_evolution}, we study the co-evolution of galaxies with their supermassive black holes. The top panel shows the M$_\star$-M$_\textrm{BH}$ relation for galaxies and their black holes at $z=2$ in cosmosTNG. Contours and black points show the distribution of black hole masses at fixed galaxy stellar mass, while the black dashed line shows the median relation. Stellar masses are summed within a $2.0$\,arcsec diameter aperture. We compare against two observational samples: the gray line shows the median relation with its $1-\sigma$ band from~\citet{Tanaka24}, while the gray markers with error bars show individual observed high-z galaxies at $z>4$ with JWST/NIRSpec from~\citet{Harikane23}.

Recent observations of high redshift SMBHs have suggested they may be `overmassive' with respect to the $z=0$ expectation, given their host masses, constraining possibly seeding scenarios \citep{Maiolino24,Bogdan24}. The lower panel of Figure~\ref{fig:smbh_gal_evolution} therefore quantifies the redshift evolution of the deviation of black hole masses from the local M$_\star$-M$_\textrm{BH}$ relation, expressed by the relative offset $\Delta\log\left(\textrm{M}_\textrm{BH}/\textrm{M}_\odot\right)$:
$$\Delta\log\left(\textrm{M}_\textrm{BH}/\textrm{M}_\odot\right)=\log(\textrm{M}_\textrm{BH}/\textrm{M}_\odot)-\alpha_\mathrm{local}\log\left(\textrm{M}_\star/\textrm{M}_\odot\right) - \beta_\mathrm{local}$$
We use the local relation based on~\citet{Haring04} and~\citet{Bennert11} with $\alpha_\mathrm{local}=0.97$ and $\beta_\mathrm{local}=-2.48$~\citep{Tanaka24}. We compute the deviation $\Delta\log\left(\textrm{M}_\textrm{BH}/\textrm{M}_\odot\right)$ for all galaxies with M$_\star>10^{11.0}$\msun and show the median relation as the green line. Shaded regions show the range containing the central $68$\%, $95\%$, and all galaxies.
We find a positive median deviation, offset from the local relation by $\sim 0.5$\, dex between $z=2.0-3.5$, suggesting that massive galaxies at this epoch indeed host overmassive SMBHs.
The median of the offset decreases towards higher redshift, and drops below the local relation at $z\sim 4.5$. Observations from~\citet{Tanaka24} at $z\sim 2$ are broadly consistent with cosmosTNG\@. The observational sample from~\citet{Sun24} indicates no significant redshift evolution between $z\sim 2-4$, compatible with the simulation. In contrast, observations at $z>4$ from~\citet{Harikane23} indicate much larger, positive deviations from the distribution in cosmosTNG\@. However, we find that this difference is driven largely by the (lower) stellar masses of observed host galaxies. The blue line shows the maximum $\Delta\log\left(\textrm{M}_\textrm{BH}/\textrm{M}_\odot\right)$ offset in cosmosTNG when we adopt a lower galaxy mass cutoff of M$_\star>10^{8.5}$\msun. In this case, deviations of up to $2$\,dex from the local median relation exist across the entire redshift range. We conclude that systems with mass ratios similar to the~\citet{Harikane23} sample can be found in cosmosTNG, although they may represent outliers to the general population. This suggests the need for a more careful analysis of selection functions, in order to assess whether observed black hole masses in galaxies at $z \sim 4$ are compatible with cosmosTNG, and TNG model simulations in general.

\subsection{Run Variations within Protocluster Regions}
\label{sec:protoclusterregion}

\begin{figure*}
    \includegraphics[width=\textwidth]{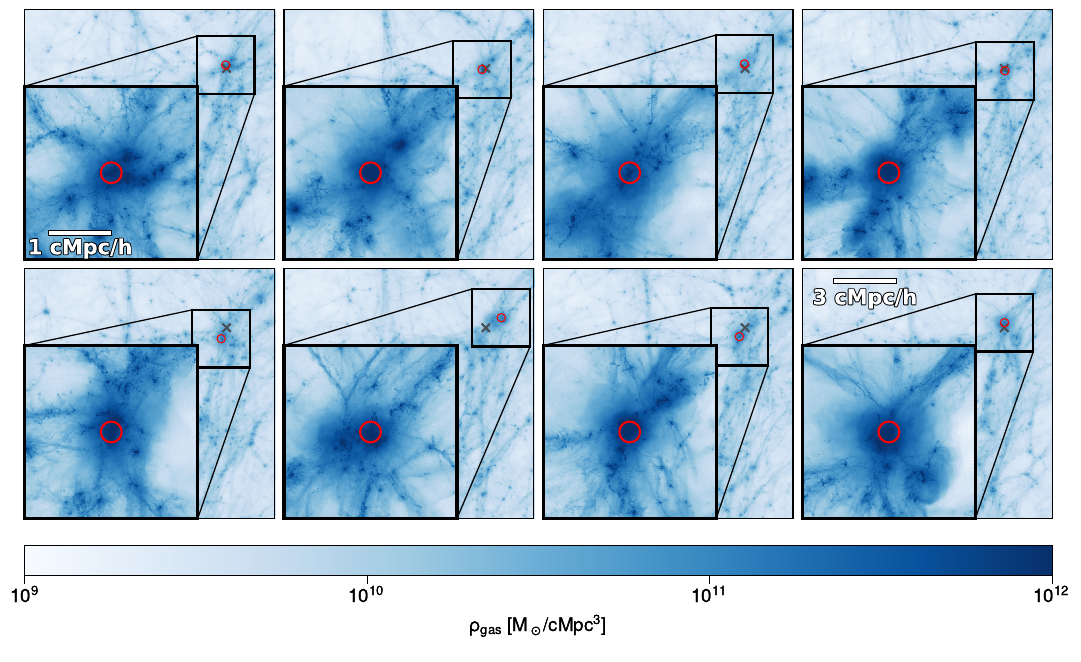}
    \includegraphics[width=\textwidth]{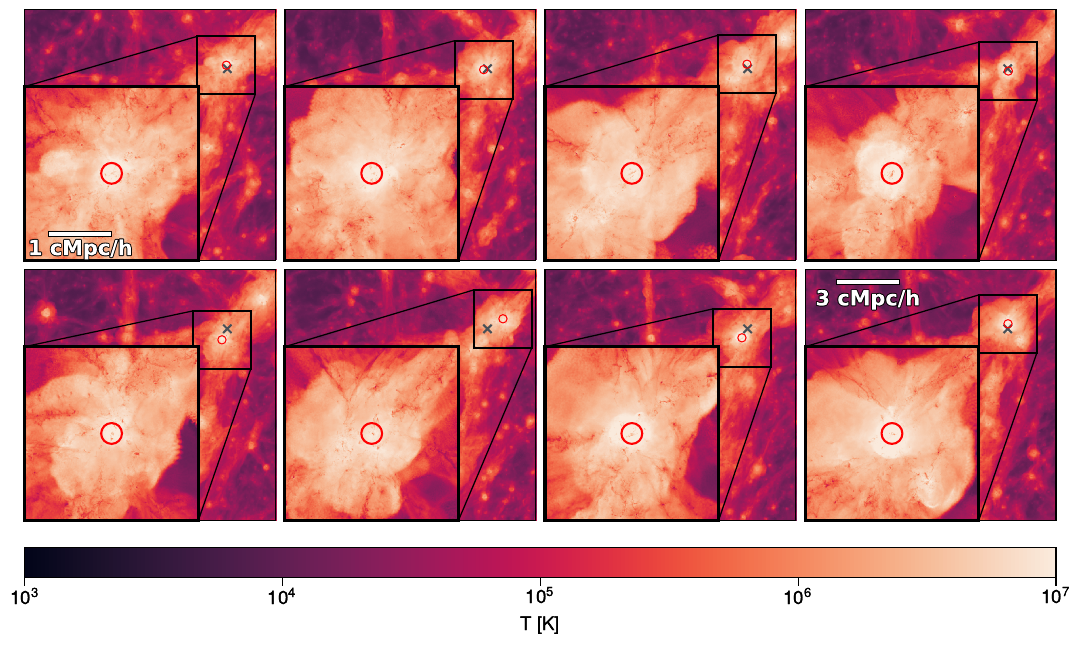}
    \caption{Gas density and temperature projection at $z=2$ for the same comoving protocluster region across all 8 variation runs in the R.A.-Dec. plane with a projection depth of $30$\,cMpc/h. In the zoom-in inset, we center on the most massive halo in the protocluster region with a projection depth of $3$\,cMpc/h. Red circles indicate the halo center, and gray crosses the mean position across variation runs. \textbf{Top}: Gas Density. \textbf{Bottom}: Temperature. We find similar density and temperature fields on the largest spatial scales. While the most massive halo appears nearly unchanged across variations, other halos seem to change substantially.}
    \label{fig:protoregion}
\end{figure*}

In Figure~\ref{fig:protoregion} we focus in on one of the unique regions in the cosmosTNG volume, and its properties across our eight variation runs: the zFIRE protocluster ($z=2.11$). In particular, we show projections of gas density and temperature in the R.A.-Dec. plane. The inset zooms in on the most massive halo in the field. 

From the density projection, we clearly see the similarities of large-scale structure across the variations. This provides a clear sanity check that the constrained modes of the initial conditions are fixing the large-scale structure in cosmosTNG\@. Even more striking, we can clearly identify the same massive halo across all the runs. In fact, its spatial position varies by less than $200$\,pkpc from the mean, indicating an excellent level of coherency between the variation runs. Nevertheless, the halo mass varies by $\sim 50\%$ across the 8 variation runs (Table~\ref{tab:variations}), which is a larger box-to-box variance compared with most of the other statistics. However, this is unsurprising because this halo is the most non-linear object in the volume that renders its properties difficult to reconstruct precisely. For less massive objects, the level of persistence between variation runs further decreases, although the overall orientation of the matter distribution with the embedding cosmic web still agrees well.
In the temperature projection, we see a similar large-scale resemblance across variations. Here, the zoom-in panels reveal large, Mpc-sized hot outflows arising from the AGN activity of the most massive halo, i.e. the protocluster BCG. However, there are sizable differences in the physical properties of this AGN-driven outflow, including in its spatial extent. 
In subsequent work, we will study whether these large-scale outflows can explain the `missing' COSTCO-I protocluster, which is detected through its galaxy overdensity but not in its Lyman-$\alpha$ forest signature (\citealt{Dong23}, \citealt{Dong24}; see also \citealt{Lee16}).

\begin{figure*}
    \includegraphics[width=\columnwidth]{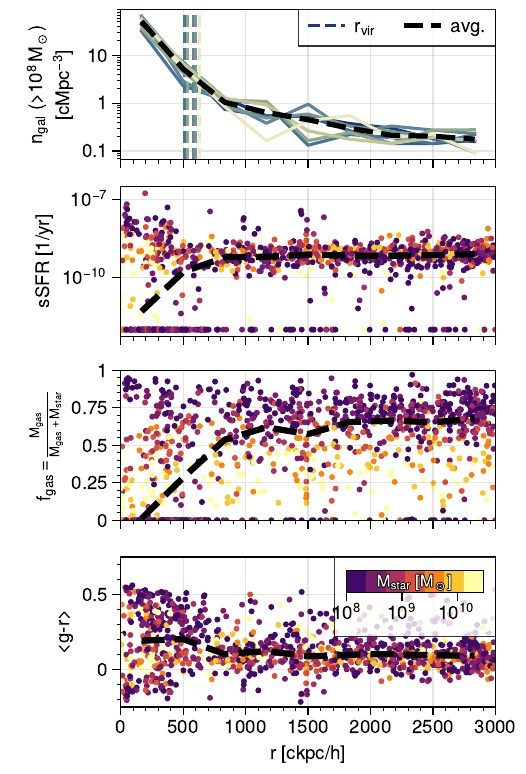}
    \includegraphics[width=\columnwidth]{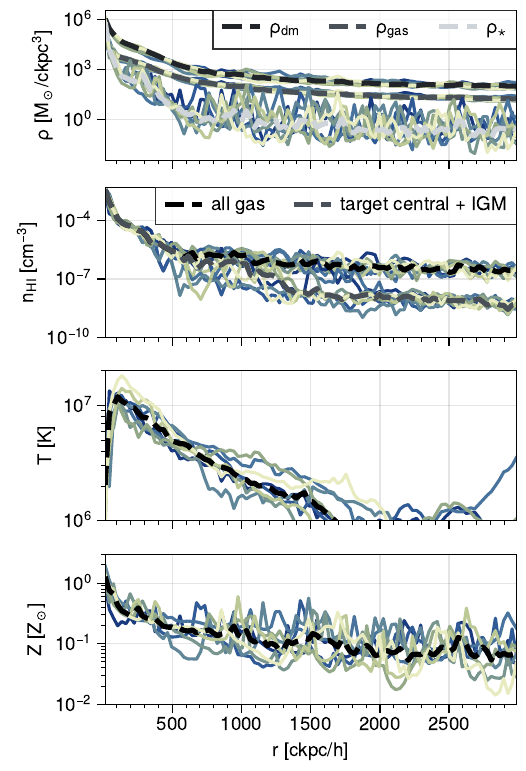}
    \caption{Radial profiles of galaxies, galaxy properties, and physical properties of the underlying dark matter, gas, and stars, centered on the most massive halo in the zFIRE protocluster region. \textbf{Left:} We show the median distribution of the number of surrounding galaxies, plus their sSFR, gas fraction, and $\langle \mathrm{g}-\mathrm{r}\rangle$ color, including galaxies with M$_\star$>$10^{8}$\msun. The virial radii of the main zFIRE halo are indicated by the vertical dashed lines. \textbf{Right:} In the top panel, we show the radial density profiles for dark matter, gas and stars. In the lower three panels, we show the neutral hydrogen density, temperature and metallicity radial profiles respectively. Each colored line shows another variation run, while the black line shows their median.}
    \label{fig:protoregion_radialprofiles}
\end{figure*}

To quantify the physical structure of the zFIRE protocluster region, Figure~\ref{fig:protoregion_radialprofiles} shows several radial profiles centered on the most massive halo. On the left, we plot the galaxy number density profile, i.e. the number of satellite and/or nearby galaxies, as well as three properties of those systems: sSFR, gas fraction $f_\mathrm{gas}$, and $\langle \mathrm{g}-\mathrm{r}\rangle$ color (intrinsic, dust-free). In all cases, as a function of the three-dimensional distance to the most massive halo in the zFIRE protocluster region. On the right, we show radial profiles of dark matter, gas and stellar density, as well as the neutral hydrogen density, gas temperature, and gas-phase metallicity.

In the left panels of Figure~\ref{fig:protoregion_radialprofiles}, only galaxies with stellar masses above $10^8$\msun are included. We find strong trends of their specific star formation rate and gas fraction as a function of distance from the protocluster center, clearly signaling the impact of environmental effects, even beyond the virial radius of the central protocluster halo (vertical dashed lines). Galaxies within this boundary are, on average, less star-forming, often quenched, and significantly gas depleted. Although the median sSFR and f$_\mathrm{gas}$ decreases towards the protocluster center, we simultaneously find an increased number of gas-rich and rapidly star-forming galaxies, typically with low masses $10^8$-$10^9$\msun. This overall change in star-formation is just weakly imprinted in the $\langle \mathrm{g}-\mathrm{r}\rangle$ color with an $0.1$\,mag increase from the overall median of $\langle \mathrm{g}-\mathrm{r}\rangle\sim 0.1$\,mag. Such outliers may indicate peculiar environmental effects such as the enhancement of SFR due to compression of the star-forming interstellar medium \citep{Vulcani18,Goeller23}. A detailed comparison of how galaxies evolve in the particular environments of the zFIRE and Hyperion protoclusters, and how this evolution differs from typical environmental effects at the same overdensity, will be explored in future work.

The right panels of Figure~\ref{fig:protoregion_radialprofiles} reveal that the large-scale mass distribution of dark matter and gas is similar between the eight variation runs, out to several Mpc. The stellar density field shows the most variation, due to the varying positions of individual high-stellar mass systems. Variation in the neutral hydrogen density, as well as gas temperature, reflect the strong impact of baryonic feedback effects, including stellar and AGN-driven outflows, as well as the impact of the AGN radiation field. Of particular interest, these differences will be reflected in the observables of Lyman-alpha, including halo-scale and cosmic web-scale emission \citep{Byrohl21,Byrohl23}, as well as absorption in the forest as used in the reconstruction of the underlying density field for the constrained initial conditions. Both topics, in the particular environments of the COSMOS field, will be the subject of future work.

\subsection{Galaxy Clustering}

\begin{figure}
    \includegraphics[width=\columnwidth]{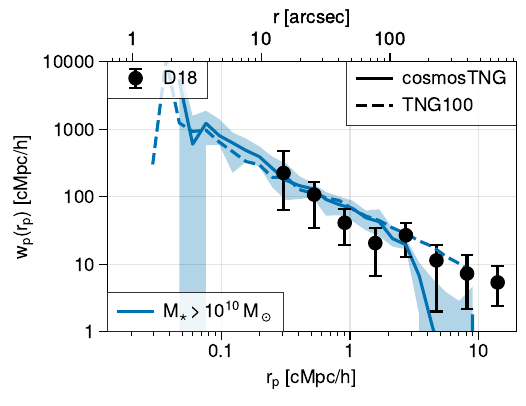}
    \caption{Projected two-point correlation function for galaxies with stellar masses above $10^{10.0}$\msun at $z=2.73$ in cosmosTNG and TNG100. We show observational data with the same mass cut for $\langle z \rangle=2.8$ from~\citet{Durkalec18}.}
    \label{fig:tpcf}
\end{figure}

We close by extending our analysis beyond one-point statistics, to better understand the spatial distribution of galaxies and their clustering in cosmosTNG. Figure~\ref{fig:tpcf} shows the projected two-point correlation function of $\omega(\mathrm{r})$ for cosmosTNG and TNG100 at $z\sim 2.73$ for galaxies with stellar masses above $10^{10.0}$\msun. We compute the projected correlation function with a line-of-sight integration limit of $\pi_\mathrm{max}=20$\,cMpc/h as
\begin{equation}
\omega_\mathrm{p}(r_\mathrm{p})=2\int_0^{\pi_\mathrm{max}} \xi\left(r_\mathrm{p}, \pi\right)\mathrm{d}\pi,
\end{equation}
using the Landy-Szalay estimator for the underlying correlation function $\xi\left(r_\mathrm{p}, \pi\right)$~\citep{Landy93}.

There is a mild clustering excess in cosmosTNG compared to TNG100, particularly at smaller scales. The clustering signal in cosmosTNG quickly drops off above $r_p\geq 3$cMpc/h due to the limited plane of the sky extent of the simulated volume. For comparison, we show observations for the clustering of galaxies above the same mass threshold at $\langle \mathrm{z}\rangle\sim 2.8$ from~\citet{Durkalec18}. Both TNG simulations are consistent with the observed clustering signal. This improves over earlier comparisons of cluster in COSMOS with theoretical models \citep[e.g.][]{McCracken07}. Due to our limited redshift depth, we cannot compare to projected clustering measurements based on photometric redshifts. However, in addition to the prominent zFIRE and Hyperion protocluster regions, clustering algorithms have recently enabled robust catalogs of galaxy assemblies down to the group mass scale, and out to $z \gtrsim 2$ \citep{Toni24}, which we can use to further explore the impact of environment and environmental effects on galaxy evolution in the COSMOS field.

\section{Discussion and Future Prospects}

\subsection{Constrained simulations at Cosmic Noon}

Galaxy properties are shaped by their large-scale environment, and interactions in dense environments are an important influence for the galaxy population. Constrained cosmological simulations are an important tool to understand and test our galaxy formation models in this aspect. In the local universe at $z \sim 0$ they directly probe the assembly of observed clusters and their galaxies. At high redshift, galaxies evolve even more rapidly, and have more frequent interactions, and constrained simulations similarly enable us to study the vigorous star-formation and merger activity at Cosmic Noon.

Constrained cosmological simulations are complementary to conventional cosmological galaxy formation simulations using random realizations for the initial conditions. First, they enable a more meaningful and direct comparison with observational surveys in the reconstructed volumes, given that the large-scale structure matches, minimizing biases from cosmic variance. Second, constrained simulations allow us to investigate the biases and non-linear relations between different observed tracers. Specifically, we have knowledge of both the diffuse gas distribution (\Lya forest) and bright star-forming galaxies (via photometric and spectroscopic surveys) in the COSMOS field. As a result, we can assess how well the constrained simulation agrees with different observed tracers.

The resulting (dis)agreement reflects how well a given field, observable, or galaxy population traces the underlying matter density field \citep[e.g.][]{Momose22}. Simultaneously, it provides a somewhat new vantage point onto the evaluation of the fidelity of the underlying galaxy formation model. These two effects are partially degenerate, but one could vary the information i.e. tracer(s) used during the reconstruction procedure in order to differentiate their relative roles. Such an approach may be particularly insightful at high redshifts, contrasting absorption from HI in the \Lya forest versus galaxy population observations, where the relationship and connections between the two may be non-trivial.

Constrained simulations at the epoch of $z \sim 2$ Cosmic Noon provide further practical advantages: notably, the availability of Lyman-alpha forest tomography for mapping the diffuse gas large scale structure~\citep{Lee14a}. Computationally, the compute time requirement is also only 20\% of an analogous simulation run all the way to $z=0$. In the present work, this has enabled us to run the eight variations of the same volume, in order to better marginalize over unconstrained small-scale structure.

\subsection{Uniqueness of the cosmosTNG target region}
\label{sec:uniqueness}

\begin{figure}
    \includegraphics[width=0.95\columnwidth]{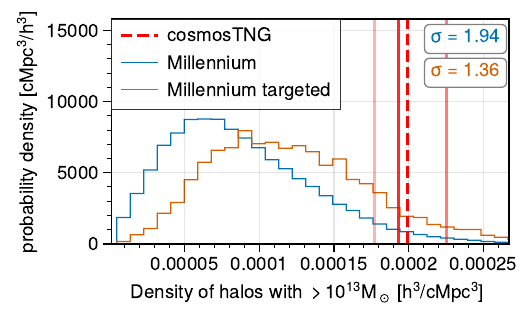}
    \caption{Assessment of the rarity of the cosmosTNG volume. Specifically, we randomly place cosmosTNG sub-volumes within the Millennium-I simulation, and measure the number density of halos with M$_{h}$>$10^{13}$\msun. The probability density for many random sub-volumes (blue) is contrasted against the same but for volumes with a M$_{h}$>$10^{13}$\msun halo at the relative position of the zFIRE cluster within cosmosTNG (orange). The vertical red lines compare to the number density measured within the cosmosTNG variation runs, and the dashed line shows their mean.}
    \label{fig:peculiarity}
\end{figure}

CLAMATO and the cosmosTNG volume target an overdense region, as indicated by the excess of massive halos in the HMF (see Appendix Figure~\ref{fig:resolutionstudy}). Part of this can already be seen in the initial conditions (Figure~\ref{fig:power_constrained}) showing an excess in power on constrained large spatial scales between $k=0.2$ and $1$\,h/cMpc. This excess could be partly physical, related to the specific constrained field, and partly related to the TARDIS pipeline~\citep[see e.g. Figure 4 in][]{Horowitz21b}.

To better understand the region, we quantify the uniqueness of the cosmosTNG subvolume in terms of the number density of massive halos with M$_{h}$>$10^{13}$\msun. We compare the mean value of different realizations to the probability density function inferred from the much larger, dark matter only cosmological simulation Millennium~\citep{Springel05}. The result is shown in Figure~\ref{fig:peculiarity}. This probability distribution is computed by randomly sampling subvolumes of Millennium (boxsize $500$cMpc$^{3}$/h$^{3}$) with the cosmosTNG geometry and computing the number of massive halos each contains (blue distribution). As we have selected the cosmosTNG subvolume given the presence of the zFIRE protocluster, we also repeat this procedure under the constraint of one such massive halo is present at the same relative position as in cosmosTNG (orange distribution). This is done by first selecting massive halos in Millennium and centering the masking volume by the offset between zFIRE and the cosmosTNG subvolume center.

We find standard deviations of $\sigma=1.9$ ($\sigma=1.4$) for the cosmosTNG subvolume when compared against a random (targeted) realization, demonstrating that the cosmosTNG volume is a considerably overdense environment. Such environment is not present in TNG100. Note that the standard deviation is approximate given the differing cosmologies of cosmosTNG and Millennium.

\subsection{Galaxy Evolution Modulated by Large-Scale Structure}

Using the TNG galaxy model and comparing to a fiducial realization in TNG100, we find significant changes in many summary statistics presented in Section~\ref{sec:results}. These include: an early peaking of the cosmic star-formation rate density, an enhancement in the stellar mass function at the high-mass end, a stellar-mass shift of the quenched fraction, and a mild clustering excess.

These findings are qualitatively consistent with the presence of more pronounced large-scale overdensities (Section~\ref{sec:uniqueness}), most prominently the zFIRE galaxy protocluster at $z\approx2.1$. In particular, observations suggest that large-scale overdensities modulate the normalization of the stellar mass function, and in doing so over-proportionally enhance the number density of massive galaxies~\citep{Daikuhara24, Forrest24}. Curiously, we also find a shift of the quenched fraction as a function of stellar mass towards higher galaxy masses, corresponding to a mildly lower quenched fraction at fixed mass (Figure~\ref{fig:ssfr_vs_mstar}) when compared to the TNG100 simulation. This is one of the most striking differences in cosmosTNG with respect to other TNG model simulations. It suggests that the particular large-scale environment of the COSMOS field has a strong and observable impact on the history and/or frequency of galaxy quenching.

In order to better understand the robustness and consistency of such differences, we ran eight variations of the same simulation volume. All have the large-scale density field fixed by the TARDIS constraints, but randomly inject power for $k\geq1$\,h/cMpc where observational constraints are missing. Overall, we find only limited variance in outcomes caused by the random modes for various galaxy summary statistics such as the SMF, SFRD and clustering signal. Only at high masses do significant variations start to appear due to low number statistics of the underlying massive halos and their more non-linear nature. Variation-to-variation differences are smaller than the difference of cosmosTNG to TNG100. This suggests that the large-scale density field (i.e. modes at $k>1$\,h/cMpc) have a strong and overwhelming impact on galaxy population statistics and scaling relations at this level.

On the scale of individual protocluster regions (Section~\ref{sec:protoclusterregion}), however, substantial variations arise. While the density profiles are similar, protohalo masses (Table~\ref{tab:variations}) and their temperature and metallicity profiles (Figure~\ref{fig:protoregion}) often vary by a factor of $2$ or more. This indicates the lack of small-scale power constraints as well as constraints on the formation history of these regions.

The TARDIS pipeline imposes no strong cosmological constraints on the field reconstruction, even though its forward model evolves the fields at fixed cosmology~\citep{Horowitz21, Horowitz21b}.
The resulting power excess on large scales (Figure~\ref{fig:power_constrained}) might in parts indicate the uniqueness of the simulated COSMOS subvolume as an overdense field, as well as an underestimated bias of the hydrogen Lyman-$\alpha$ optical depth. Furthermore, the TARDIS pipeline itself has shown to introduce excess power on intermediate constrained scales~\citep[see][]{Horowitz21b}.
This potential uniqueness moreover implies that caution should be exercised when interpreting the observed $2\lesssim z \lesssim 2.5$ galaxy properties from surveys that have significant overlap with the CLAMATO field (Figure~\ref{fig:cosmosfield}), such as COSMOS-Web \citep{Casey23}.
There are also possible ramifications toward cosmological surveys due to the heavy dependence on the COSMOS field for calibrating photometric redshifts \citep[e.g.,][]{Masters17} --- further analysis might be required on the full $\sim 1$ deg$^2$ COSMOS field to check whether overdensities might be biasing these calibrations.

\subsection{Reconstruction Methods and Future Directions}
\label{sec:discuss_reconstruction}

The initial conditions, particularly Fourier modes on resolved, reconstructed scales $k\leq 1$\,h/Mpc, can strongly affect the observed properties of frequently-studied massive galaxies. We now discuss shortcomings and future directions for the reconstruction pipeline and use in cosmological galaxy formation simulations.

The cosmosTNG simulations are run as a downstream step following reconstruction using the TARDIS inference scheme applied on joint Lyman-$\alpha$ forest and galaxy redshift data sets. Ultimately, we would want to integrate galaxy formation simulations within the reconstruction pipeline. This would enable an `end-to-end' forward model, at the level of galaxy observables, or even at the field level. However, key challenges exist in this regard. First, the current likelihood maximization method makes use of the differentiability of its forward model. Despite recent progress~\citep{Li24}, this remains out-of-reach for complex galaxy formation simulation models such as TNG\@. Furthermore, the computational costs would certainly be prohibitive, without model surrogate or emulation techniques. In the interim, future work can aim to increase consistency between the expensive downstream galaxy formation simulations and the cheap reconstruction forward model.

A major improvement to the reconstruction scheme will stem from an improved treatment and coupling of the two complementary tracers used in this study, namely diffuse gas traced by the \Lya forest and bright star-forming galaxy populations. Both are subject to different biases and modeling uncertainties. In the TARDIS-based reconstruction applied here, bright galaxies from the zCOSMOS survey and \Lya absorption from the CLAMATO survey were used simultaneously, even though the former generally influences only the overall amplitude of the reconstruction and not the detailed filamentary structure~\citep{Horowitz21b}.

As such, we are particularly susceptible to systematics in the mapping of observed \Lya forest to the underlying density field. In the reconstruction, the TARDIS scheme assumes the FGPA, which becomes increasingly inconsistent with the hydrodynamical result at redshift $z\sim 2$~\citep{Kooistra22}. Recent observations indicate that in some cases galaxies and \Lya forest do not just not trace each other, but are even anti-correlated. For example, \citet{Dong23} found that some \Lya forest reconstructed underdensities host overdensities in galaxies, while \citet{Newman22} found environments with the opposite relation, where overdense neutral hydrogen regions are underdense in galaxies.

Future explorations with cosmosTNG will revisit the \Lya forest in cosmosTNG, studying its deviations from the FGPA and potential protocluster heating around massive objects, to be parameterized in the forward model~\citep{Horowitz21b}. As we have shown, when we can faithfully map the \Lya forest to the underlying matter field, the large-scale overdensities encoded in the initial conditions lead to the spatially coherent emergence of massive protocluster regions below the reconstruction scale (see Figure~\ref{fig:protoregion}). This might indicate an underrepresented relative weight of massive galaxies in the reconstruction pipeline.

Current and upcoming data from surveys such as JWST PRIMER, COSMOS-Web, Blue Jay and COSMOS-3D~\citep{Dunlop21,Casey23,Belli24,Kakiichi24} in the field will not only allow a better comparison with simulation outcomes, but will add substantial constraining potential above the current cutoff scale of $k\gtrsim 1\,$\,h/cMpc. For example, the protohalo studied in Section~\ref{sec:protoclusterregion} is currently only covered by few zCOSMOS galaxies. Hence, much of the information provided by the distribution of shown zFIRE galaxies is currently missing, which can be included along with other survey data in future reconstructions. The reconstruction can further benefit from additional tracers (i.e. orthogonal probes) other than \Lya forest and galaxies, such as the cosmic web \Lya emission~\citep{Bacon21,Byrohl23,Martin23}.

We have also only simulated a partial region of the reconstructed CLAMATO field (Figure~\ref{fig:cosmosTNG450}) in this work. Extensions of cosmosTNG toward other parts of the field, e.g., the large cosmic void at $z\approx 2.35$ (\citet{Krolewski2018}; see top panels of Figure~\ref{fig:cosmosTNG450}) could reveal interesting astrophysics.

\section{Conclusions}
\label{sec:conclusions}

In this introductory paper we present the new cosmosTNG simulation suite. This is a set of cosmological galaxy formation simulations run with the IllustrisTNG model and with `constrained' initial conditions inferred from the galaxy and gas distribution within the COSMOS field, evolved to Cosmic Noon at $z=2$. We have compared our results to the TNG100 simulation with conventional random initial conditions, as well as to observations across a variety of galaxy properties, including star-formation rates, quenching, supermassive black hole masses, environmental effects, and clustering.

Our main findings are:

\begin{itemize}
\item The reconstructed initial conditions, evolved with the TNG galaxy formation model, gives broadly consistent results with the TNG100 cosmological galaxy formation simulation. Simultaneously, the distribution of simulated galaxies at $z \sim 2$ is roughly and qualitatively consistent with the observed ($\alpha$, $\delta$, $z$) distribution of galaxies from spectroscopic surveys.
\item At the high-mass end, cosmosTNG shows key differences with respect to a random realization. Most notably, it has an enhanced number of halos and galaxies, i.e. more high-stellar-mass galaxies than expected ($\sim 2\times$ for $M_\star\geq 10^{11}$\,M$_\odot$), qualitatively consistent with excess power in the constrained volume. These systems have overall increased star-formation activity, and a lower quenched fraction in comparison to TNG100 ($\Delta_{\text{q. frac.}}\sim20$\,\% at $M_\star=10^{11}$\,M$_\odot$).
\item The abundance of massive galaxies at $2 < z < 5$ is in reasonable agreement with observational inferences. The abundance of massive quiescent galaxies is low, potentially lower than observations suggest, although the volume of cosmosTNG is too limited to draw stronger conclusions. 
\item cosmosTNG shows an intriguing signal in the relationship between supermassive black hole (SMBH) and galaxy stellar mass, as suggested by recent JWST observations. In particular, with respect to the local $z=0$ relation, cosmosTNG galaxies host overmassive black holes, by $\sim 0.5$ dex at $2 \lesssim z \lesssim 4$. While this average difference largely vanishes at $z > 4$, we show that lower mass galaxies with $M_\star \sim 10^9$\msun can have SMBHs overmassive by up to 2\,dex up to $z=5$.
  \item The properties of galaxies, and of the underlying dark matter, gas, and stellar matter fields around the zFIRE protocluster hint towards unique imprints of environmental effects on galaxy evolution in this region, while all showing a qualitiatively consistent large-scale structure on the reconstruction scale ($k=1$\,h/cMpc).
\end{itemize}

Our results show notable differences for galaxy properties at the high-mass end $>10^{11}$\msun, demonstrating the need for caution for assessing the simulation's realism or their astrophysical/cosmological implications when contrasted with observed high-redshift survey volumes.
Future analyses will focus on further comparing the observed properties of galaxies specifically within the COSMOS region to simulated outcomes, enabling a new theoretical vantage point on galaxy formation and environmental effects on galaxy evolution at Cosmic Noon.

\section*{Acknowledgments}

All of our post-processing and analysis of the simulations makes use the~\texttt{scida} analysis library~\citep{Byrohl24}.

CB and DN acknowledge funding from the Deutsche Forschungsgemeinschaft (DFG) through an Emmy Noether Research Group (grant number NE 2441/1-1). We also thank the Hector Fellow Academy for their funding support. This work was further supported by the Deutsche Forschungsgemeinschaft (DFG, German Research Foundation) under Germany's Excellence Strategy EXC 2181/1 - 390900948 (the Heidelberg STRUCTURES Excellence Cluster). This research was supported in part by grant NSF PHY-1748958 to the Kavli Institute for Theoretical Physics (KITP). The authors gratefully acknowledge the scientific support and HPC resources provided by the Erlangen National High Performance Computing Center (NHR@FAU) of the Friedrich-Alexander-Universität Erlangen-Nürnberg (FAU) under the NHR project a103bc. NHR funding is provided by federal and Bavarian state authorities. NHR@FAU hardware is partially funded by the German Research Foundation (DFG) – 440719683. Additional simulations and analysis were carried out on the Vera machine of the Max Planck Institute for Astronomy (MPIA) and systems at the Max Planck Computing and Data Facility (MPCDF).
KGL acknowledges support from JSPS Kakenhi Grant JP24H00241.
Kavli IPMU is supported by the World Premier International Research Center Initiative (WPI), MEXT, Japan.

\section*{Data Availability}

The cosmosTNG simulations will be made publicly available and accessible at \url{www.tng-project.org/data} \citep{Nelson19} in the near future. Data directly related to this publication is available on request from the corresponding author.

\bibliographystyle{aa}
\bibliography{references}

\clearpage
\begin{appendix}

\section{Resolution Convergence}

\begin{figure*}
    \includegraphics[width=\textwidth]{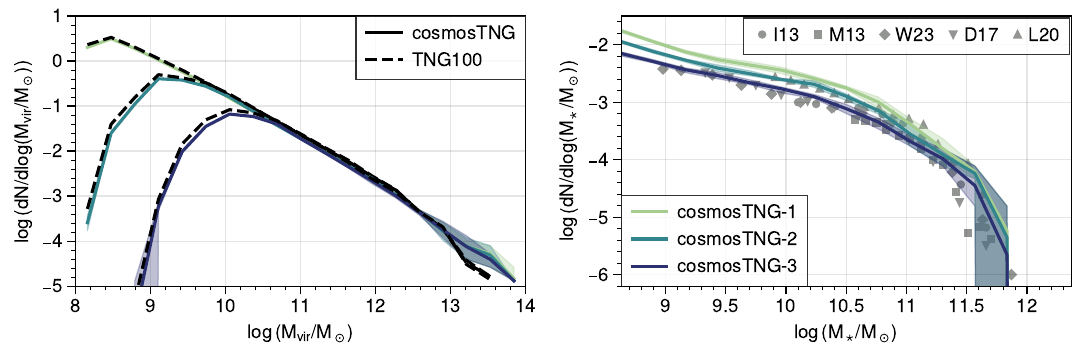}
    \caption{Resolution convergence for cosmosTNG. \textbf{Left:} Halo mass function (HMF) for three different resolutions. The dashed lines show the corresponding TNG100 resolution run. \textbf{Right:} Stellar mass function for three different resolutions and gray markers for observational data. The HMF converges down to a characteristic turnover at the low-mass end below $10^{8.5}$\,M$_\odot$ for cosmosTNG-1. The cosmosTNG HMF matches well with its respective TNG100 pendant except for $M_\mathrm{vir}>10^{13}$\msun where we find an excess of massive halos in cosmosTNG. The stellar mass function generally appears to retain shape at increased resolution, but systematically increasing by up to $0.2$\,dex per resolution level at fixed stellar mass.}
    \label{fig:resolutionstudy}
\end{figure*}

In Figure~\ref{fig:resolutionstudy}, we show the resolution convergence of cosmosTNG in the halo and stellar mass function. Resolution levels are indicated by the suffix number as defined in Table~\ref{tab:resolution}.
For the halo mass function, we find good convergence for masses down to a resolution-dependent turnover at the low-mass end, which lies below $10^{8.5}$\,M$_\odot$ for cosmosTNG-1, and proportionally larger by a factor of 8 according to the mass resolution for low resolution runs. In contrast, the stellar mass function shows a systematic increase in the number of galaxies at fixed stellar mass with increasing resolution, with a difference of up to $0.2$\,dex per resolution level, in agreement with previous findings for the employed star formation model~\citep{Springel03} in cosmological volumes~\citep{Genel14}.

\end{appendix}

\end{document}